\documentclass[preprint,12pt]{elsarticle}

\journal{Journal of Applied Geophysics}
\usepackage[english]{babel}
\usepackage[utf8]{inputenc}
\usepackage[T1]{fontenc}
\usepackage{amssymb,amsmath}
\usepackage{graphicx}
\usepackage{caption}
\usepackage{appendix}
\usepackage[colorlinks]{hyperref}   
\usepackage[figuresonly,nomarkers,nolists]{endfloat} 

\begin{document}

\begin{frontmatter}
\title{Offset-continuation-trajectory stacking based on common-reflection-point kinematics for five-dimensional prestack dataset regularization and enhancement}

\author[1,2,3]{Tiago Coimbra\corref{cor1}}
\ead{tgo.coimbra@gmail.com}
\author[1,4]{Rodrigo Bloot}
\author[1,2,3]{Caian Benedicto}
\author[1,2,3]{Nicholas Torres Okita}
\author[1,2,3]{Jorge Henrique Faccipieri Junior}
\address[1]{High-Performance Geophysics Laboratory (HPG Lab)}
\address[2]{Centro de Estudos de Energia e Petróleo (CEPETRO)}
\address[3]{Universidade Estadual de Campinas (UNICAMP), Campinas, Brazil}
\address[4]{Federal University of Latin-American Integration (UNILA), Foz do Iguaçu, Brazil}
\cortext[cor1]{Corresponding author}
\begin{abstract}

Prestack seismic data regularization and enhancement are critical steps for reliable imaging and inversion, particularly in five-dimensional (5D) dataset geometries affected by irregular sampling, noise contamination, and incomplete spatial coverage. These limitations often degrade event continuity and compromise the physical consistency of conventional interpolation methods. This study introduces a physics-informed framework for 5D prestack dataset reconstruction based on a multi-parameter common-reflection-point (CRP) traveltime stacking operator. The proposed offset-continuation-trajectory (OCT) operator derives coherent stacking trajectories from wavefront propagation, isochronous surface geometry, specular reflection, and diffraction kinematics. All kinematic parameters are estimated directly from the data through a global coevolutionary optimization strategy. The method reconstructs missing traces and enhances spatial continuity by stacking seismic events along physically consistent traveltime surfaces, preserving both reflection and diffraction kinematics. Applications to synthetic and field datasets demonstrate improved signal-to-noise ratio, enhanced structural continuity, and reliable recovery of unrecorded amplitudes without introducing artificial events. The results indicate that incorporating physically constrained traveltime models into the regularization process provides a robust, geologically consistent alternative to purely mathematical interpolation techniques, thereby improving data fidelity for subsequent imaging and quantitative interpretation.

\end{abstract}

\begin{keyword}
5D interpolation \sep CRP stacking \sep wavefront kinematics
\end{keyword}

\end{frontmatter}

\newpage

\section{Introduction}

Irregularities in the acquisition geometry of seismic datasets, caused by acquisition constraints, pose significant challenges for various processing and imaging algorithms. These limitations arise from factors such as economic, environmental, logistical, or technical limitations, leading to inadequate coverage or even gaps in critical areas \citep{vermeer:2012, stolt:2013}. Moreover, during seismic acquisition, environmental noise or external sources such as wind, rain, ocean waves, natural seismic activity, and even human activities close to the acquisition area can contaminate the data, along with electronic noise introduced by the components of the seismic acquisition system, including amplifiers, cables, transducers, and other devices, which generate significant interference in the seismic signals. To mitigate these issues, interpolation, regularization, and enhancement techniques applied directly to prestack datasets have become an active and growing area of research. A wide range of approaches has been proposed in the literature, encompassing both model-driven and data-driven methodologies. For example, we can find representative examples and recent developments in \cite{ma2025irregularly,park2025improving,liu:2004,kaur2019seismic,wang2019deep,mikhailiuk2018deep}. 

In this context, seismic acquisition noise can pose a significant challenge for multidimensional interpolation, degrading the quality of the interpolated data. Strategies to mitigate the impact of noise in prestack seismic data include filtering, denoising, robust statistical techniques, and coherent stacking. Filtering techniques, such as frequency-wavenumber ({\it F-K}) and median filters, are commonly applied to preserve horizontal coherence while attenuating incoherent noise across the gathers \citep{yilmaz:2001}. Denoising approaches, such as curvelet transforms, have demonstrated efficiency in separating signal from noise due to their ability to sparsely and directionally represent seismic events \citep{herrmann:2008a}. Robust statistical models, such as those using principal component analysis (PCA) or robust singular value decomposition (SVD), can suppress coherent noise while preserving signal attributes \citep{trickett:2008}. Coherent stacking, particularly phase-consistent or semblance-based stacking, is widely employed to enhance primary reflections and attenuate incoherent energy, even before velocity analysis or imaging \citep{neidell:1971}. These strategies collectively aim to improve the signal-to-noise ratio (SNR) while preserving the integrity of physical events for subsequent processing steps such as velocity analysis, migration, and inversion. However, some of these techniques may not be sufficient to improve the SNR or may even introduce undesirable information in the worst-case scenario.

Prestack dataset interpolation algorithms aim to fill in the gaps by estimating data values at unsampled or unrecorded locations, using information from nearby observed points \citep{trad:2007, spitz:1991}. After interpolation is performed, the estimated values populate missing positions, generating a more spatially regular seismic dataset. This process inherently involves assumptions and approximations, which may rely purely on numerical and mathematical models \citep{abma:2006} or incorporate physical principles, such as wavefront propagation \citep{coimbra:2016}. The effectiveness of interpolation depends on the quality and spatial distribution of the input data, as well as the appropriateness of the chosen method \citep{herrmann:2008b}. A robust interpolation algorithm should preserve the kinematics of both reflection and diffraction events, accurately modeling wavefront behavior as it traverses heterogeneous media and encounters structural discontinuities. Ideally, it should also maintain amplitude relationships to ensure faithful reconstruction of the seismic energy \citep{liu:2004}. Therefore, careful dataset analysis and awareness of method-specific limitations are crucial for mitigating uncertainties in the interpolation process.

Several approaches have been proposed to interpolate and regularize seismic data, including Fourier-based techniques \citep{xu:2005}, curvelet-domain thresholding \citep{herrmann:2008b}, and rank-reduction methods in higher dimensions \citep{kreimer:2012}. However, many of these methods rely heavily on purely mathematical assumptions and often struggle to maintain accurate wavefield kinematics, especially in complex geological settings. More recent efforts have sought to incorporate physical models into the interpolation process, such as offset-continuation operators \citep{fomel:2003} and wave-equation-based interpolators \citep{biondi:2006}. However, they still may require regular acquisition geometry to function effectively.

To address these challenges, we propose a five-dimensional (5D) regularization and interpolation strategy based on a multi-parametric traveltime-stacking operator rooted in the common-reflection-point (CRP) concept. In contrast to conventional interpolation techniques, our method employs a physics-informed stacking framework that estimates reflection responses along kinematically consistent traveltime trajectories. The stacking operators are defined by a set of kinematic parameters, which are calculated directly from the data through global optimization using a coevolutionary algorithm \citep{Ribeiro:2023:ENS}. This strategy enables the generation of an interpolated dataset that adheres to the physical principles of seismic wave propagation, thereby preserving the curvature and amplitude behavior of reflection events, even in irregular or sparse acquisition geometries.

The proposed operator extends the traditional offset continuation trajectory (OCT) stacking methodology, as presented by \cite{coimbra:2016}, to five dimensions, thereby enhancing its applicability to modern 5D datasets obtained from 3D complex acquisition layouts. By formulating the stacking operator for each target amplitude grid point, the method allows adaptive interpolation in both midpoint and offset domains, effectively recovering missing traces while maintaining physical consistency.

To validate the method, we apply it to both synthetic and real field datasets. The results demonstrate that the new operator improves signal continuity, reduces acquisition-related noise, and enhances the SNR. These outcomes demonstrate that incorporating physical wavefront models into the interpolation process yields more robust and geologically consistent results than purely mathematical methods.

In summary, we address the challenges associated with irregularly sampled 5D prestack seismic datasets by introducing a physics-informed regularization and interpolation framework based on CRP kinematics. The proposed approach extends the OCT methodology to the whole 5D prestack domain, allowing the reconstruction and enhancement of seismic data along physically consistent traveltime trajectories. By directly estimating all kinematic parameters from the data using a global coevolutionary optimization strategy, the method preserves both reflection and diffraction behavior while mitigating acquisition-related gaps and noise contamination. The results obtained from synthetic and real datasets indicate that the proposed framework achieves its intended objectives, providing improved signal-to-noise ratio, enhanced event continuity, and geologically consistent reconstructions. Although convergence stability may be limited in extremely sparse or highly heterogeneous regions, the method demonstrates robust performance under realistic acquisition conditions, representing a significant advancement over purely mathematical or numerical interpolation techniques.

Finally, we organize the paper as follows. In the next section, we introduce the main concepts and theoretical foundations underlying the proposed interpolation and regularization framework, and we derive the key mathematical expressions and technical details. We then present and discuss the CRP operator and its associated 5D OCT traveltime formulation. In the subsequent section, we apply the method to both synthetic and field datasets to illustrate its performance. Furthermore, we provide the discussion, conclusions, acknowledgments, and references.

\section{Methodology}

This section presents the methodological framework underlying the proposed approach. We begin with a theoretical description of the wave propagation model and the construction of isochronous and reflection traveltime surfaces, providing the physical and mathematical background for the operator. Then, we explore the properties of diffraction traveltime, highlighting its role in characterizing non-specular events and its relationship with the reflection traveltime surface. Next, we introduce the offset continuation trajectory, which captures the kinematic signature of reflection points in the CRP domain as a function of the half-offset. Finally, we examine the intrinsic consistency of the OCT traveltime formulation, validating its coherence under physical and geometrical constraints relevant to seismic data regularization and enhancement.

\subsection{Theoretical description}

To begin our discussion about the method, we need to define a two-dimensional (2D) geometric surface, denoted $\Lambda$, also known as the seismic acquisition or measurement surface, to describe our structural model. In our framework, $\Lambda$ is a horizontal plane embedded in a three-dimensional (3D) space, $(x,y,z)$, located at $z = z_0$. This surface represents the area where seismic instruments, such as sources and receivers, are deployed to record waves generated by controlled sources. The choice of a horizontal acquisition plane simplifies the initial analysis of wavefront propagation, providing a consistent reference for our physical and mathematical framework. As a result, the seismic dataset comprises an array of traces recorded during surveys, designed according to specific acquisition parameters.

To analyze the kinematic behavior of seismic wavefronts using traveltime information extracted from 5D prestack seismic data, we introduce $\tau({\bf x}_0;\Tilde{\bf x}_D)$ as the function that measures the traveltime from point $\Tilde{\bf x}_0=({\bf x}_0,z_0)$ on $\Lambda$ to a subsurface point $\Tilde{\bf x}_D$. Observe that in our notation, $\Tilde{\bf x}_D=(x_D,y_D,z_D)$ is a 3D point at depth, and ${\bf x}_0 = (x_0,y_0)$ is a 2D point lying on the measurement surface. In our analysis, we parameterize the position of the source and receiver in $\Lambda$ by ${\bf m}$ and ${\bf h}$, where the expressions (${\bf m}-{\bf h}$) and (${\bf m}+{\bf h}$) correspond to the coordinates of the source and receiver, respectively. Here, ${\bf m}$ and ${\bf h}$ are the midpoint 2D-point and the half-offset 2D-vector between the source-receiver position.

The traveltime function $\tau_E$ measures a three-point path (source-event-receiver) and is defined as
\begin{equation}
\tau_E({\bf m},{\bf h};\Tilde{\bf x}) = \tau({\bf m}-{\bf h};\Tilde{\bf x}) + \tau({\bf m}+{\bf h};\Tilde{\bf x})\, ,
\label{eq:tauE}
\end{equation}
where $\tau$ represents the exact one-way traveltime from the measurement acquisition surface to some depth point. In other words, this traveltime of a three-point event measures the time it takes for a seismic wave to travel between three specific points. The wavefront starts at the seismic source on the measurement surface, travels to the event point underground, and then returns to the receiver on the measurement surface. Therefore, given a depth point $\Tilde{\bf x}_D$, we can define the traveltime $t_0$ located in a reference trace to the source-receiver pair parameterized by ${\bf m}_0$ and ${\bf h}_0$ as
\begin{equation}
t_0 = \tau_E({\bf m}_0,{\bf h}_0;\Tilde{\bf x}_D)\, .
\end{equation}
Note that we currently define the traveltime $t_0$ as the response time of a three-point event, but we have not yet specified the nature of the event occurring at the second point (i.e., depth point).

\begin{figure}
    \centering
    \includegraphics[width=0.8\linewidth]{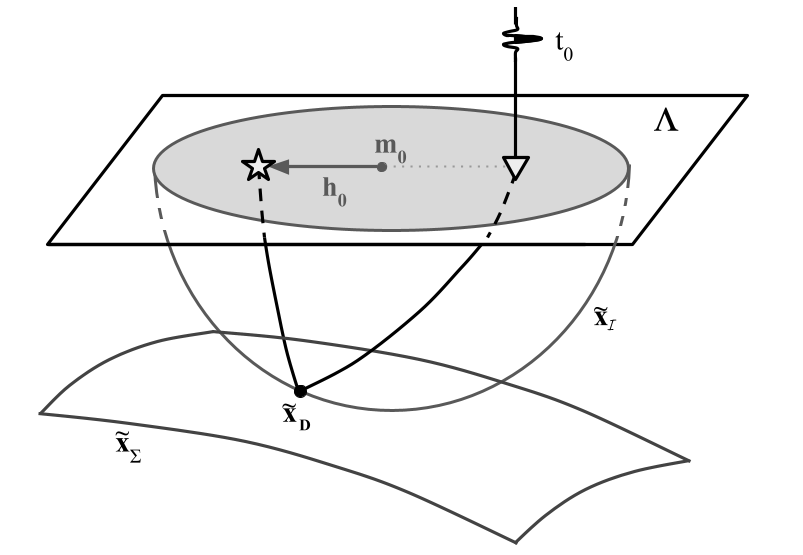}
    \caption{Schematic representation of the acquisition geometry. The measurement plane is denoted by $\Lambda$, with the source and receiver represented by the star and the inverted triangle, respectively. The isochrone surface, generated by $({\bf m}_0,{\bf h}_0,t_0)$, and the reflecting interface are denoted by $\tilde{\bf x}_{\cal I}$ and $\tilde{\bf x}_\Sigma$, respectively. The point $\tilde{\bf x}_D$ indicates the tangency point between the two surfaces.}
    \label{fig:isocrona}
\end{figure}

Given the point $({\bf m}_0,{\bf h}_0)$ and the traveltime $t_0$, we define 
\begin{equation}
\Tilde{\bf x}_{\cal I} = \Tilde{\bf x}_{\cal I}(\boldsymbol{\gamma})
\end{equation}
as the surface parameterized by $\boldsymbol{\gamma}=(\gamma_1,\gamma_2)$ in $\Gamma$, such that there exists a point $\boldsymbol{\gamma}_D$ in $\Gamma$ for which $\Tilde{\bf x}_D = \Tilde{\bf x}(\boldsymbol{\gamma}_D)$ with $\tau_E({\bf m}_0,{\bf h}_0;\Tilde{\bf x}_{\cal I}) = t_0$ for all $\boldsymbol{\gamma}$ in $\Gamma$ as illustrated in Figure~\ref{fig:isocrona}. Therefore, $\Tilde{\bf x}_{\cal I}$ is what we define as an isochronous surface referring to $({\bf m}_0,{\bf h}_0,t_0)$, which refers to a surface underground where the traveltime of a three-point path, that is, the time it takes for the wave to travel from the seismic source to the receiver passed by any point at $\Tilde{\bf x}_{\cal I}$, is constant. In other words, all points on the surface $\Tilde{\bf x}_{\cal I}$ have the same arrival time of the seismic wave for the source-receiver pair on the measurement surface defined by ${\bf m}_0$ and ${\bf h}_0$.

Once we have defined our isochronous surface, we can demonstrate its relationship to the reflecting interface via specular reflection. Specular reflection links the isochronous surface to the reflection surface, since both will have the same point of contact and the same normal. Consequently, these properties ensure that the two surfaces are tangent at the point $\Tilde{\bf x}_D$. To address this analysis, we employ the following parameterization
\begin{equation}
\Tilde{\bf x}_{\Sigma} = \Tilde{\bf x}_{\Sigma}(\boldsymbol{\gamma})
\end{equation}
as the surface parameterized by $\boldsymbol{\gamma}$ in ${\Gamma}$. This parameterization defines the reflection surface, which represents all possible reflection points that contribute to the seismic response observed at the surface as described in Figure~\ref{fig:isocrona}.

Additionally, we define two reparameterized traveltime functions that enable us to express the two-way (three-point path) traveltime as a function of the acquisition geometry and subsurface parameters 
\begin{equation}
\left\{
\begin{array}{ccc}
\tau_{\cal I}({\bf m},{\bf h},\boldsymbol{\gamma}) &=& \tau_E({\bf m},{\bf h};\Tilde{\bf x}_{\cal I})\, , \\
\tau_\Sigma({\bf m},{\bf h},\boldsymbol{\gamma}) &=& \tau_E({\bf m},{\bf h};\Tilde{\bf x}_{\Sigma})\, ,
\end{array}
\right.
\label{eq:tE}
\end{equation}
where $\tau_{\cal I}$ denotes the traveltime associated with the isochronous surface, and $\tau_\Sigma$ corresponds to the traveltime evaluated on the reflecting interface surface. Both are functions of the midpoint ${\bf m}$, half-offset ${\bf h}$, and the parameterization vector $\boldsymbol{\gamma}$, providing a framework for analyzing the seismic response in terms of reflection geometry and traveltime kinematics.

Based on the proposed hypothesis, in which the isochronous surface serves as a first-order osculating surface (i.e., it is tangent at a common point) to the reflecting interface at $\Tilde{\bf x}_D$, we can express the corresponding geometric condition through the following mathematical equalities
\begin{equation}
\left\{
\begin{array}{ccc}
\Tilde{\bf x}_{\Sigma}(\boldsymbol{\gamma}_D) &=& \Tilde{\bf x}_{\cal I}(\boldsymbol{\gamma}_D)\, , \\
\left.\dfrac{\partial\Tilde{\bf x}_{\Sigma}}{\partial\boldsymbol{\gamma}}\right|_{\boldsymbol{\gamma}_D} &=& \left.\dfrac{\partial\Tilde{\bf x}_{\cal I}}{\partial\boldsymbol{\gamma}}\right|_{\boldsymbol{\gamma}_D}\, .
\end{array}
\right.
\label{eq:samex}
\end{equation}
It is also important to note that this condition may occur at multiple points, and in some cases, the reflecting interface may even coincide entirely with the isochronous surface. However, for our analysis, we focus on a single point of tangency.

Therefore, for a specific depth point $\Tilde{\bf x}_D$ combined with the conditions in equation~(\ref{eq:samex}), we define $\Lambda_R$ as the set of all midpoint–offset pairs $({\bf m}, {\bf h})$ that satisfy the following stationary condition
\begin{equation}
\left.\frac{\partial \tau_\Sigma}{\partial \boldsymbol{\gamma}}\right|_{\boldsymbol{\gamma}_D}=\left.\frac{\partial \tau_{\cal I}}{\partial \boldsymbol{\gamma}}\right|_{\boldsymbol{\gamma}_D} = {\bf 0}\, .
\label{eq:specular}
\end{equation}
In other words, on the isochronous and reflection surfaces parameterized by ${\boldsymbol{\gamma}}_{\cal I}$ and ${\boldsymbol{\gamma}}_{\Sigma}$, respectively, there exist points $({\bf m}, {\bf h})$ in $\Lambda_R$ such that both $\tau_\Sigma$ and $\tau_{\cal I}$ represent the specular reflection traveltimes for the same depth point, corresponding to the value 
\begin{equation}
\boldsymbol{\gamma}_{\cal I}({\bf m}, {\bf h}) = \boldsymbol{\gamma}_{\Sigma}({\bf m}, {\bf h}) = \boldsymbol{\gamma}_D    
\end{equation}
that satisfies equation~(\ref{eq:specular}). Therefore, taking $({\bf m},{\bf h})$ in $\Lambda_R$, we can define the reflection traveltime on the isochronous surface as 
\begin{equation}
\tau_{\cal I}({\bf m},{\bf h},\boldsymbol{\gamma}_{\cal I}({\bf m},{\bf h})) = \tau_\Sigma({\bf m},{\bf h},\boldsymbol{\gamma}_\Sigma({\bf m},{\bf h}))\, . 
\label{eq:tite}
\end{equation}

Despite the continuity of the surface $\Tilde{\bf x}_\Sigma$, situations can arise in which $\Tilde{\bf x}_\Sigma$ is not smooth, i.e., the first derivative concerning $\boldsymbol{\gamma}$ does not exist at the point $\boldsymbol{\gamma}_D$ for some points in $\Lambda_R$. In such cases, a tip diffraction occurs at that location. In other words, this condition corresponds to a point where seismic waves scatter from the edges of geological structures, such as faults or other geological subsurface discontinuities \citep{klem:2003edge}. Therefore, this situation indicates that the condition specified by equation~(\ref{eq:tite}) is not fulfilled. Thus, at locations on $\Lambda_R$ where the derivative with respect to $\boldsymbol{\gamma}$ is undefined, indicating non-smooth behavior of the surface, we assign $\boldsymbol{\gamma}_\Sigma({\bf m},{\bf h}) = \boldsymbol{\gamma}_D$ to denote the corresponding tip diffraction point. Synthesizing the concepts discussed above, we formulate the reflection traveltime as
\begin{equation}
\tau_R({\bf m},{\bf h}) = \tau_\Sigma({\bf m},{\bf h},\boldsymbol{\gamma}_{\Sigma}({\bf m},{\bf h}))\, . 
\end{equation}

Once we have defined the reflection traveltime at point $\Tilde{\bf x}_D$, we can now define the traveltime for a diffraction response from a single scatter point at $\Tilde{\bf x}_D$. Based on equation~(\ref{eq:tE}), we set the diffraction traveltime for such situation as
\begin{equation}
\tau_D({\bf m},{\bf h}) = \tau_E({\bf m},{\bf h};\Tilde{\bf x}_D)\, .
\label{eq:tauD}
\end{equation}

\begin{figure}
    \centering
    \includegraphics[width=0.8\linewidth]{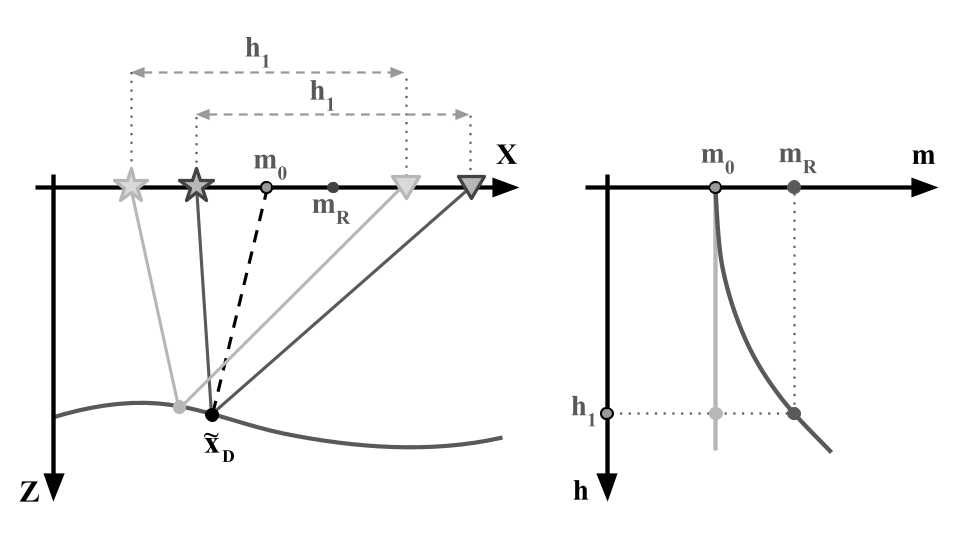}
    \caption{Variation of midpoint position as a function of half-offset for a 2D acquisition. A comparison is made with a conventional source-receiver pair in a CMP configuration for the same fixed half-offset (light gray) where the reflection point over the refletor deviates from $\Tilde{\bf x}_D$.}
    \label{fig:crp2d}
\end{figure}

The variation of the midpoint as a function of the half-offset relative to a consistent reflection point at depth describes the projected CRP surface. This surface captures how the midpoint coordinates change with varying half-offsets, providing trajectories based on half-offset parameters running over $\Lambda_R$. In order to make it, we define the set $\Lambda_\text{CRP}$ as the points $({\bf m},{\bf h})$ in $\Lambda_R$ such that 
\begin{equation}
\tau_D({\bf m},{\bf h}) = \tau_R({\bf m},{\bf h})\, .
\label{eq:CRP_def}
\end{equation}
Therefore, setting the function ${\bf m} = {\bf m}_{R}({\bf h})$ where ${\bf m}_0 = {\bf m}_{R}({\bf h}_0)$ and the surface $({\bf m}_{R}({\bf h}), {\bf h})$ belongs to $\Lambda_\text{CRP}$ defines the projected CRP trajectory which is illustrated in Figure~\ref{fig:crp2d}.

\subsection{Diffraction traveltime properties}

We now examine the diffraction traveltime defined in equation~(\ref{eq:tauD}) to investigate key kinematic properties of wavefront propagation as recorded on the acquisition surface. Specifically, we focus on identifying a stationary point located within the zero-offset (ZO) section. According to image-ray theory \citep{hubral:1977time}, there exists at least one midpoint ${\bf x}_0$ at ${\bf h}_0 = {\bf 0}$ such that
\begin{equation}
\left.\frac{\partial \tau_D}{\partial {\bf m}}\right|_{({\bf x}_0,{\bf 0})} = \left.\frac{\partial \tau_D}{\partial {\bf h}}\right|_{({\bf x}_0,{\bf 0})} = {\bf 0}\, .
\end{equation}
This point, denoted by $\Tilde{\bf x}_0=({\bf x}_0,z_0)$ and lying on the surface $\Lambda$, corresponds to a coincident source-receiver location and represents the apex of the diffraction response, where the recorded traveltime reaches the global minimum.

In the following, we consider a depth model that includes an arbitrary and unspecified velocity medium. Besides, such a medium is assumed to be sufficiently regular to permit a second-order Taylor series expansion of the traveltime $\tau_E$ in the vicinity of $({\bf x}_0,{\bf 0})$ with $\Tilde{\bf x}_D$ fixed. Based on these assumptions, we can make a hyperbolic approximation of the one-way traveltime function $\tau$, in equation~(\ref{eq:tauE}), with respect to the first variable. This procedure provides an adequate description of wavefront propagation information in the dataset domain without requiring a detailed analysis of the medium. Therefore, we can represent the diffraction-traveltime response as
\begin{eqnarray}
\tau_D({\bf m},{\bf h}) &=&  \frac{1}{2}\sqrt{\tau_0^2 + (\Delta{\bf x}_0-{\bf h})^T{\bf S}_0(\Delta{\bf x}_0-{\bf h})} \nonumber \\
 &+& \frac{1}{2}\sqrt{\tau_0^2 + (\Delta{\bf x}_0+{\bf h})^T{\bf S}_0(\Delta{\bf x}_0+{\bf h})} \nonumber \\
 &+&{\cal O}\left(\|\Delta{\bf x}_0\pm{\bf h}\|^3\right)\, ,   
 \label{eq:tauD2}
\end{eqnarray}
where $\Delta{\bf x}_0 = {\bf m}-{\bf x}_0$, $\tau_0 = \tau_E({\bf x}_0,{\bf 0};\Tilde{\bf x}_D)$, the function ${\cal O}$ represent the Big-O notation that informs the order of approximation, and the symmetric $2\times 2$ matrix
\begin{equation}
{\bf S}_0 = \tau_0\left.\frac{\partial^2 \tau_E}{\partial{\bf m}\partial {\bf m}^T}\right|_{({\bf x}_0,{\bf 0};\Tilde{\bf x}_D)} = \tau_0\left.\frac{\partial^2 \tau_E}{\partial{\bf h}\partial {\bf h}^T}\right|_{({\bf x}_0,{\bf 0};\Tilde{\bf x}_D)}\, .  
\end{equation}

In the special case of a homogeneous and isotropic medium, the expressions involved in the diffraction traveltime formulation simplify significantly due to the absence of velocity gradients and anisotropy. As demonstrated by \cite{coimbra:2019enhancement}, the higher-order terms vanish, and the second-order behavior becomes explicitly defined as
\begin{equation}
\begin{array}{ccc}
{\cal O}\left(\|\Delta{\bf x}_0\pm{\bf h}\|^3\right) &\equiv& 0\, , \\
{\bf S}_0 &=& \frac{4}{v^2}{\bf I}\, ,
\end{array}
\label{eq:tauD_v}
\end{equation}
where ${\bf I}$ denotes the $2 \times 2$ identity matrix, and $v$ represents the constant propagation velocity of the wavefront. In this context, the traveltime surface around the diffraction apex assumes a locally hyperbolic shape, and the curvature matrix ${\bf S}_0$ becomes isotropic and spatially invariant. This simplification facilitates the analytical treatment of diffraction responses and provides a useful starting point for understanding more complex heterogeneous scenarios.

\textbf{Remark 1:} Given a propagation medium based on our premises, the expression in equation~(\ref{eq:tauD2}) approximates the diffraction traveltime by accounting for perturbations around the apex. However, this formulation inherently introduces a third-order approximation error. This error arises from deviations in the wavefront geometry from the idealized spherical shape, typically assumed under the hypothesis of a homogeneous and isotropic medium. In real media, heterogeneities and anisotropic features distort the wavefront curvature, leading to higher-order perturbations in the traveltime surface. Specifically, the third-order terms capture the cumulative effect of these physical deviations, making the approximation less accurate as the offset and midpoint increase in proportion to the complexity of the medium.

\textbf{Remark 2:} In contrast, under the approximation described in equation~(\ref{eq:tauD_v}), the matrix ${\bf S}_0$ remains symmetric, positive-definite, and the traveltime exhibits a spatially invariant curvature centered at the diffraction apex. This configuration corresponds to the idealized scenario of a homogeneous and isotropic medium, where the wavefront curvature remains constant with position. Consequently, as already mentioned, the leading-order error in the traveltime approximation is limited to third-order terms in the presence of perturbations due to the heterogeneity or anisotropy of the medium. This simplification, as we see later, not only facilitates more tractable analytical modeling and efficient numerical implementation but also offers a valuable reference framework for quantifying and interpreting the effects of medium heterogeneity on the diffraction response.

\subsection{A double-square-root Ansatz for diffraction traveltime approximation}

To enable efficient computation of the traveltime operator, we adopt an Ansatz in which the traveltime is locally approximated by a hyperbolic expression centered at the image-ray arrival point on the measurement surface. This formulation preserves the essential kinematic behavior in the vicinity of that point and remains valid under the assumption of a locally elliptical wavefront. The proposed Ansatz not only simplifies analytical modeling but also provides a foundation for constructing the kinematic attributes of the OCT.

Based on the model described in equation~(\ref{eq:tauD2}), we adopt the following analytical expression to represent the diffraction traveltime response measured on the acquisition surface as
\begin{equation}
t_D({\bf m},{\bf h};{\bf x}_0) = t_u({\bf m}-{\bf h},{\bf x}_0) + t_u({\bf m}+{\bf h},{\bf x}_0)\, ,   
\end{equation}
where the following hyperbolic Ansatz is defined as
\begin{equation}
t_u({\bf u},{\bf x}_0) =  \frac{1}{2}\sqrt{\tau_0^2 + ({\bf u}-{\bf x}_0)^T{\bf S}_0({\bf u}-{\bf x}_0)}\, ,   
\label{eq:ansatz}
\end{equation}
in which ${\bf u}$ denotes a generic point on the measurement plane $\Lambda$, and ${\bf x}_0 \in \Gamma$ is a reference location that generates an image-ray path from the acquisition surface point $({\bf x}_0, z_0)$ to its corresponding depth point $\Tilde{\bf x}_{\cal I}({\bf x}_0)$. The scalar traveltime parameter $\tau_0$, associated with the image-ray path, is defined as
\begin{equation}
\tau_0 = \tau_0({\bf x}_0) = \tau_E({\bf x}_0,{\bf 0};\Tilde{\bf x}_{\cal I}({\bf x}_0))\, . 
\label{eq:tauzero}
\end{equation}

These expressions, known as the double-square-root (DSR) diffraction traveltime approximation, provide a kinematic approximation that captures the primary characteristics of wavefront propagation from a diffraction point within the subsurface. By construction, the model yields the following relation for a stationary point ${\bf x}_0$, as
\begin{equation}
\tau_D({\bf m},{\bf h}) = t_D({\bf m},{\bf h};{\bf x}_0) + {\cal O}\left(\|\Delta{\bf x}_0\pm{\bf h}\|^3\right)\, ,    
\end{equation}
where $\Delta{\bf x}_0 = {\bf m} - {\bf x}_0$. Accordingly, the third-order error term arises from distortions in the wavefront caused by variations in the medium, such as anisotropy or heterogeneity. 

Before we proceed to the deduction and analysis of the two-way traveltime associated with an image-ray path, as defined in equation~(\ref{eq:tauzero}), it is important to take into account that the partial derivatives of $\tau_0$ with respect to ${\bf x}_0$ depend on $\tau_E$. As a consequence, applying the chain rule to this expression, we obtain
\begin{equation}
\frac{\partial\tau_0}{\partial{\bf x}_0} = \frac{\partial\tau_E}{\partial{\bf x}_0} + \frac{\partial\Tilde{\bf x}_{\cal I}^T}{\partial{\bf x}_0}\frac{\partial\tau_E}{\partial\Tilde{\bf x}_{\cal I}}\, .    
\label{eq:derivadatauzero}
\end{equation}

By design, the function $\Tilde{\bf x}_{\cal I}({\bf x}_0)$ is defined such that $\Tilde{\bf x}_{\cal I}$ corresponds to the subsurface point satisfying $t_0 = \tau_E({\bf m}_0, {\bf h}_0; \Tilde{\bf x}_{\cal I})$ as ${\bf x}_0$ varies along the acquisition surface. Furthermore, as previously defined, the image-ray path connects the surface location $({\bf x}_0, z_0)$ to its corresponding image point $\Tilde{\bf x}_{\cal I}({\bf x}_0)$, where this path satisfies the stationary traveltime condition with respect to ${\bf x}_0$. Therefore, for every point ${\bf x}_0$ on the measurement surface, the partial derivative of the extended traveltime $\tau_E$ with respect to ${\bf x}_0$ vanishes, i.e.,
\begin{equation}
\frac{\partial \tau_E}{\partial {\bf x}_0} = {\bf 0}\, .
\end{equation}

According to the definition given by equation~(\ref{eq:tauzero}), the traveltime $\tau_0({\bf x}_0)$ corresponds to the isochronous response interpreted as a reflector in the time-migrated domain. Therefore, the derivative in equation~(\ref{eq:derivadatauzero}) represents the local slope of this event at ${\bf x}_0$, which we denote as
\begin{equation}
\frac{\partial \tau_0}{\partial {\bf x}_0} = {\bf d}_0\, .
\label{eq:dtau0dx0}
\end{equation}
Geometrically, the vector ${\bf d}_0$ defines the gradient of the isochronous traveltime surface at the point ${\bf x}_0$ on the measurement plane. In the context of seismic migration, this gradient is directly related to the apparent dip of the event in the migrated image, which in turn reflects the local orientation of the reflecting interface. Thus, ${\bf d}_0$ plays a central role in determining the kinematic consistency between the modeled isochronous and the true reflection geometry by the equality of their image slope property. 

Now, based on our proposed Ansatz, we compute the first derivative of the traveltime function $t_u$ with respect to a generic acquisition coordinate $\mathbf{u}$. Applying the chain rule to the expression defined in equation~(\ref{eq:ansatz}), we obtain the following relation
\begin{equation}
4t_u\,\mathbf{a}_u = \mathbf{S}_0(\mathbf{u} - \mathbf{x}_0)\, ,
\label{eq:au}
\end{equation}
where
\begin{equation}
\mathbf{a}_u = \frac{\partial t_u}{\partial \mathbf{u}}
\end{equation}
represents the slowness vector of the traveltime function with respect to $\mathbf{u}$. This expression describes the directional derivative, characterizing the rate of change of the traveltime with respect to the measurement domain.

As stated above, the position $({\bf m}_0,{\bf h}_0,t_0)$ is known. We now incorporate the local slope information at this same position. Let us define ${\bf u}$ as either the source location ${\bf s} = {\bf m} - {\bf h}$ or the receiver location ${\bf r} = {\bf m} + {\bf h}$, such that
\begin{equation}
\mathbf{a}_s = \frac{\partial t_s}{\partial \mathbf{s}} \quad \text{and} \quad
\mathbf{a}_r = \frac{\partial t_r}{\partial \mathbf{r}}\,.
\end{equation}
Consequently, the slope vectors $\mathbf{a}_{s0}$ and $\mathbf{a}_{r0}$ are also known and fixed at $({\bf m}_0,{\bf h}_0)$. According to equation~(\ref{eq:au}), there exists a set of parameters ${\bf S}_*$, ${\bf x}_*$, and $\tau_*$ such that $t_0 = t_D({\bf m}_0, {\bf h}_0; {\bf x}_*)$ and
\begin{equation}
\mathbf{a}_{s0} = \left.\frac{\partial \tau_D}{\partial \mathbf{s}}\right|_{({\bf m}_0,{\bf h}_0)} \quad \text{and} \quad
\mathbf{a}_{r0} = \left.\frac{\partial \tau_D}{\partial \mathbf{r}}\right|_{({\bf m}_0,{\bf h}_0)}\,.
\label{eq:xaster}
\end{equation}
These derivatives represent the local slopes of the diffraction traveltime with respect to the source and receiver coordinates, respectively, and are essential for determining the kinematic parameters of the underlying approximation model. Starting from the reference point $({\bf m}_0,{\bf h}_0)$, we express the approximation in a local neighborhood around this point as
\begin{equation}
\tau_D({\bf m},{\bf h}) = t_D({\bf m},{\bf h};{\bf x}_*) + {\cal O}\left(\|\Delta{\bf m}_0 \pm \Delta{\bf h}_0\|^2\right)\, ,
\end{equation}
where $\Delta{\bf m}_0 = {\bf m} - {\bf m}_0$ and $\Delta{\bf h}_0 = {\bf h} - {\bf h}_0$ denote the deviations from the reference midpoint and half-offset positions, respectively. It is important to recognize that this approximation arises from local variations in the wavefront geometry near the source–receiver pairs. Specifically, it characterizes how the wavefront curvature and two-way traveltime vary around the diffraction apex, directly influencing the kinematic behavior of the recorded events. Nonetheless, the effects of turning waves \citep[see,][]{hale:1992imaging} do not produce abnormal moveout in this context, since the adopted approach focuses on a local analysis around a reference point, where the traveltime slopes are accurately captured.

On the other hand, by taking the derivative of the adopted Ansatz concerning the position ${\bf x}_0$, we obtain an expression that combines the slope vector ${\bf d}_0$ in the time-image domain with the local slope vector $\mathbf{a}_u$, associated with either the source or receiver location. This result highlights the relationship between the traveltime gradient at the diffraction apex and the kinematic contributions from the acquisition geometry, as encoded in the Ansatz formulation. This relationship can be expressed analytically, considering the proposed Ansatz combined with the equations~(\ref{eq:dtau0dx0}) and~(\ref{eq:au}), in the
follow
\begin{equation}
4t_u \frac{\partial t_u}{\partial \mathbf{x}_0} = \tau_0 {\bf d}_0 - 4t_u\,\mathbf{a}_u\, .
\label{eq:dtudx}
\end{equation}

\subsection{Mapping to a virtual zero-offset coordinate}
\label{section:VZO}

Once we adopt the Ansatz in equation~(\ref{eq:ansatz}), we can capture the leading-order physical behavior of the traveltime response while reducing the analytical complexity of the problem. Also, we construct the adopted approximation to align with the local kinematic properties of the seismic events, particularly in the vicinity of the CRP. As such, it furnishes a physically consistent representation of the wavefront geometry under the ZO configuration, serving as a reference model for further analysis and transformation configuration.

To initiate our construction, we define a virtual ZO configuration by setting the coordinates $({\bf m}, {\bf h}) = ({\bf m}_{\text{ZO}}, {\bf 0})$, where ${\bf m}_{\text{ZO}}$ denotes the midpoint in the ZO configuration. Letting $\Delta{\bf x}_{\text{ZO}} = {\bf m}_{\text{ZO}} - {\bf x}_0$ represent the lateral distance between the ZO midpoint and the diffraction apex ${\bf x}_0$, we express the corresponding ZO traveltime, $t_{\text{ZO}} = t_D({\bf m}_{\text{ZO}}, {\bf 0};{\bf x}_0)$, as
\begin{equation}
t_{\text{ZO}}^2 = \tau_0^2 + \Delta{\bf x}_{\text{ZO}}^{\!T}\,{\bf S}_0\,\Delta{\bf x}_{\text{ZO}}\, .
\label{eq:tzo}
\end{equation}
Therefore, based on this construction, we explicitly determine the coordinates of the virtual ZO point, which corresponds to the mapped position from the apex of the diffraction event to the point where the source and receiver locations coincide on the acquisition surface. This construction is fundamental to our analysis, as it simplifies the computation of CRP trajectories. These coordinate changes play a key role in characterizing the local kinematic behavior of the wavefront and provide a reference framework for mapping and remapping across different CO configurations. Additionally, the term 'virtual ZO coordinate' refers to a theoretical coordinate in which seismic data are reinterpreted as if they had been acquired with a null offset between the source and receiver. In this context, it represents a mathematically reconstructed coordinate associated with our idealized Ansatz traveltime.

By performing the following substitution into the Ansatz formulation,
\begin{equation}
{\bf m} - {\bf x}_0 = \Delta{\bf x}_\text{ZO} + \Delta{\bf m}_\text{ZO}\, ,
\end{equation}
where $\Delta{\bf m}_\text{ZO} = {\bf m} - {\bf m}_\text{ZO}$ denotes the deviation of the midpoint ${\bf m}$ from the virtual ZO position ${\bf m}_\text{ZO}$, and using equation~(\ref{eq:tzo}), we arrive at the following expressions for the one-way traveltimes from the source and receiver positions to the diffraction apex, as
\begin{equation}
\begin{array}{ccc}
t_s &=& \dfrac{1}{2}\sqrt{t_\text{ZO}^2 + (\Delta{\bf m}_{\text{ZO}} - {\bf h})^T{\bf S}_0(\Delta{\bf m}_{\text{ZO}} - {\bf h}) + 2\Delta{\bf x}_{\text{ZO}}^T{\bf S}_0(\Delta{\bf m}_{\text{ZO}} - {\bf h})}\, , \\\\
t_r &=& \dfrac{1}{2}\sqrt{t_\text{ZO}^2 + (\Delta{\bf m}_{\text{ZO}} + {\bf h})^T{\bf S}_0(\Delta{\bf m}_{\text{ZO}} + {\bf h}) + 2\Delta{\bf x}_{\text{ZO}}^T{\bf S}_0(\Delta{\bf m}_{\text{ZO}} + {\bf h})}\, .
\end{array}
\label{eq:tstr}
\end{equation}
These expressions incorporate both the local curvature of the wavefront, through ${\bf S}_0$, and the geometric deviation from the virtual ZO apex. The terms involving $\Delta{\bf x}_\text{ZO}$ account for the offset between the diffraction apex and the center of the wavefront approximation, effectively correcting the hyperbolic moveout model for arbitrary source–receiver pairs.

Based on equation (\ref{eq:au}), particularized for the ZO situation, we obtain
\begin{equation}
t_\text{ZO}{\bf a}_\text{ZO} = {\bf S}_0\Delta{\bf x}_\text{ZO}\, ,
\label{eq:azoSxzo}
\end{equation}
which expresses the direct relationship between the slope vector ${\bf a}_\text{ZO}$ of the traveltime surface in the ZO section, i.e., the traveltime derivative in the midpoint direction at $({\bf m}_\text{ZO},{\bf 0})$, and the displacement ${\bf m}_\text{ZO}$ of the normal-incidence point concerning the reference position ${\bf x}_0$. Therefore, this relationship is fundamental, as it establishes how the geometry of the seismic event in the ZO domain is linked to its curvature and the relative position of the source and receiver at the coincidence point, serving as the basis for the accurate estimation of kinematic parameters and the construction of mapping operators.

Since $t_D = t_s + t_r$, in the specular situation, we have the following expression
\begin{equation}
    \frac{\partial t_D}{\partial \mathbf{x}_0} = {\bf 0} \, .
    \label{eq:dtddx0}
\end{equation}
Therefore,
\begin{equation}
    \tau_0 \mathbf{d}_0 = t_\text{ZO} \mathbf{a}_{\text{ZO}} = \mathbf{S}_0 \, \Delta\mathbf{x}_\text{ZO} \, .
    \label{eq:taud0tzoazo}
\end{equation}
Moreover, from the specular condition, we obtain
\begin{equation}
    \big( \Delta\mathbf{m}_\text{ZO} - \mathbf{h} \big) t_r 
    = - \big( \Delta\mathbf{m}_\text{ZO} + \mathbf{h} \big) t_s ,
    \label{eq:paralelismo}
\end{equation}
and, by isolating $\Delta\mathbf{m}_{\mathrm{ZO}}$, we arrive at the relation
\begin{equation}
    \Delta\mathbf{m}_\text{ZO} 
    = \mathbf{m}_\text{CRP}(\mathbf{h}) - \mathbf{m}_\text{ZO} 
    = \left( \frac{t_r - t_s}{t_r + t_s} \right) \mathbf{h} ,
    \label{eq:fund1}
\end{equation}
which can be used to establish a direct relationship between $\Delta \mathbf{m}_\text{ZO}$ and $\mathbf{h}$. In \cite{Coimbra:2016:TM}, the authors introduced an alignment condition that significantly simplifies the expression for the diffraction traveltime $t_D$, as given in equation~(\ref{eq:fund1}). In other words, this condition implies that the vectors $\Delta{\bf m}_{ZO}$ and ${\bf h}$ are parallel. Mathematically, this alignment is equivalent to the condition
\begin{equation}
\Delta{\bf m}_\text{ZO}{\bf h}^T = {\bf h}\Delta{\bf m}_\text{ZO}^T\, ,
\label{eq:alinhamento}
\end{equation}
which ensures that all three points lie along a common straight line. Specifically, equation~(\ref{eq:paralelismo}) becomes considerably more tractable if one assumes that the positions of the source ($\Delta{\bf m}_{ZO} - {\bf h}$), receiver ($\Delta{\bf m}_\text{ZO} + {\bf h}$), and ${\bf m}_\text{ZO}$ are collinear. This assumption facilitates the derivation of simplified kinematic expressions and enhances the tractability of further analytical developments, as demonstrated by \cite{Coimbra:2016:TM}.

By isolating $\Delta{\bf m}_\text{ZO}$ in equation~(\ref{eq:paralelismo}) and using the alignment condition given by equation~(\ref{eq:alinhamento}), we obtain  
\begin{equation}
\Delta{\bf x}_\text{ZO}^T {\bf S}_0 {\bf h} 
= \frac{t_\text{ZO}^2\, \Delta{\bf m}_\text{ZO}^T {\bf h}}
       {\|{\bf h}\|^2 - \|\Delta{\bf m}_\text{ZO}\|^2}\, .
\end{equation}
Next, substituting the expression from equation~(\ref{eq:azoSxzo}), we arrive at the following quadratic system, as
\begin{equation}
{\bf h}^T {\bf a}_\text{ZO} \left( \|{\bf h}\|^2 - \|\Delta{\bf m}_\text{ZO}\|^2 \right)
= t_\text{ZO}\, {\bf h}^T \Delta{\bf m}_\text{ZO} \, .
\end{equation}
Using the alignment condition for the system solver, we have the physical solution
\begin{equation}
\Delta{\bf m}_\text{ZO} = \xi_\text{ZO}({\bf h})\, {\bf h} \, ,
\label{eq:xizo}
\end{equation}
where the scalar function $\xi_\text{ZO}({\bf h})$ is given by
\begin{equation}
\xi_\text{ZO}({\bf h}) =
\frac{2\, {\bf a}_\text{ZO}^T {\bf h}}
     {t_\text{ZO} + \sqrt{t_\text{ZO}^2 + \left[ 2\, {\bf a}_\text{ZO}^T {\bf h} \right]^2}} \, .
\label{eq:xizoh}     
\end{equation}
This relation explicitly shows that $\Delta{\bf m}_\text{ZO}$ is collinear with the half-offset vector ${\bf h}$, and the proportionality factor $\xi_\text{ZO}({\bf h})$ depends on both the geometry of the configuration and the parameters associated with the ZO point.

By substituting equation (\ref{eq:xizo}) into the traveltime relations given in (\ref{eq:tstr}), and performing the same algebraic manipulations as in \cite{Coimbra:2016:TM}, we obtain the following expression for the DSR traveltime in the CRP configuration as
\begin{equation}
t_D\left({\bf m}_\text{CRP}({\bf h}),{\bf h};{\bf x}_0\right) 
= \sqrt{\,{\bf h}^T{\bf S}_0{\bf h} + 
\left(\frac{1}{1 - [\xi_{ZO}({\bf h})]^2}\right)t_\text{ZO}^2}\, .
\label{eq:tCRP2}
\end{equation}
Furthermore, isolating the traveltime $t_{ZO}$ in terms of a kind of NMO-corrected quantity $t_N$ in equation~(\ref{eq:tCRP2}). As a result of such manipulation, it yields
\begin{equation}
t_\text{ZO}^2 = \left(1 - [\xi_{ZO}({\bf h})]^2\right)t_N({\bf h})^2 \, ,
\label{eq:tzodmo}
\end{equation}
where
\begin{equation}
t_N^2 = t_D^2 - {\bf h}^T{\bf S}_0{\bf h}\, .
\label{eq:tN2}
\end{equation}
It is worth noting that the term $\left(1 - [\xi_{ZO}({\bf h})]^2\right)$, appearing in equation~(\ref{eq:tzodmo}), plays the role of a local correction factor. More specifically, this multiplicative factor can be interpreted as analogous to a dip moveout (DMO) correction, since it accounts for the kinematic deviation introduced by reflector dips relative to the ZO case \citep{Hale:1984dmo}. In this sense, it provides a local adjustment that modifies the traveltime expression, ensuring consistency with the geometry of dipping events.

\subsection{The slope continuation approach}

Slope continuation is a mathematical operator that describes how the local slope of reflectors in the time-space domain of seismic events changes as the offset varies. The main idea is that, similar to offset continuation, presented in \cite{coimbra:2016}, which predicts the position of an event as the source-receiver distance changes, slope continuation predicts how the traveltime derivative with respect to the midpoint coordinate (i.e., the event slope) evolves when the offset varies along the CRP trajectory. In practice, given an event with a specific slope, the equation provides its new expected slope value as a function of the offset parameter. This enables the interpolation, reconstruction, or alignment of events across midpoints. Such a tool is handy in multiparameter stacking methods, as it expands the coherence domain, improves the signal-to-noise ratio, and facilitates the reliable estimation of kinematic attributes.

By equating equations (\ref{eq:fund1}) and (\ref{eq:xizo}), and then substituting the result into equation (\ref{eq:tzodmo}), we obtain the following relation
\begin{equation}
\frac{t_D^2 t_{\text{ZO}}^2}{t_N^2} = 4t_s t_r\, .
\label{eq:relation}
\end{equation}
The last expression reveals a fundamental link between Ansatz traveltime $t_D$, virtual ZO traveltime $t_{\text{ZO}}$, and NMO-like traveltime $t_N$, in terms of contributions to the source and receiver traveltimes $t_s$ and $t_r$, respectively. This relation highlights the multiparametric consistency of the CRP-based operator and provides an analytical bridge between offset-dependent kinematics and the underlying ZO geometry.

From equation (\ref{eq:dtudx}), together with the specular reflection condition given in equation (\ref{eq:dtddx0}), and considering ${\bf a} = {\bf a}_r + {\bf a}_s$, we obtain the following relation as
\begin{equation}
t_D \tau_0 \, {\bf d}_0 = 4t_s  t_r  {\bf a} \, .
\label{eq:relationd0a}
\end{equation}
This expression links the slope in the midpoint direction represented by the combined vector ${\bf a}$ in any half-offset configuration on the CRP trajectory, and the reference slope vector ${\bf d}_0$. It provides an explicit formulation of the specular reflection geometry in terms of traveltime slopes, thereby reinforcing the slope-direction-fitting character of the CRP-based operator, as already mentioned.

Finally, by combining equations (\ref{eq:relationd0a}), (\ref{eq:relation}), and (\ref{eq:taud0tzoazo}), yields
\begin{equation}
{\bf a}_\text{ZO} = \frac{t_D t_\text{ZO}}{t_{N}^2}{\bf a}\, . 
\label{eq:azoah}
\end{equation}

\subsection{Offset continuation trajectory}

We design the OCT traveltime operator to closely fit the CRP traveltime surface at the initial measurement point, ensuring precise stacking in that neighborhood. This alignment, in an osculatory sense, enables the OCT operator to accurately represent the behavior of the CRP traveltime surface at this point. By achieving this fit, the operator provides an exact solution or, at the very least, a second-order approximation that accurately captures the essential features of wavefront propagation. This approximation is grounded in an Ansatz based on equation~(\ref{eq:tauD2}), which provides a physical and mathematical framework for understanding the underlying principles. Section \ref{section:VZO} further elaborates on this derivation and the conditions under which the approximation holds, exploring the theoretical basis and practical implications in greater detail.

Since the matrix ${\bf S}_0$ is symmetric and positive definite, we can write
\begin{equation}
\frac{4\|{\bf h}_0\|^2}{V_0^2} = {\bf h}_0^T {\bf S}_0 {\bf h}_0 \, ,
\label{eq:VhSh}
\end{equation}
where $V_0$ denotes an effective average velocity associated with that position in the domain. This relation emphasizes that the quadratic form ${\bf S}_0$ can be interpreted as a curvature matrix in a time and space domain, acting as a squared slowness matrix. In this sense, the symmetry and positive definiteness of ${\bf S}_0$ guarantee that the energy of the quadratic form is always positive. At the same time, it ensures that the value of $V_0$ defines a consistent, physically meaningful average velocity over the domain. Therefore, this simplification reduces the number of degrees of freedom without significantly sacrificing fitness adjustments.

By setting the values at the reference position in equation (\ref{eq:azoah}), we obtain
\begin{equation}
{\bf a}_\text{ZO} = \frac{t_0 \, t_\text{ZO}}{t_{N_0}^2} {\bf a}_0 \, ,
\label{eq:azo_ref}
\end{equation}
where, according to equation (\ref{eq:VhSh}), the NMO-like traveltime at the reference position satisfies
\begin{equation}
t_{N_0}^2 = t_0^2 - \frac{4 \|{\bf h}_0\|^2}{V_0^2} \, .
\label{eq:tn0}
\end{equation}
By substituting equation (\ref{eq:azo_ref}) into (\ref{eq:xizoh}), we obtain the following closed-form scalar expression for the deviation of the midpoint, given by
\begin{equation}
\xi_0({\bf h}) =
\frac{2 t_0 \, {\bf a}_0^T {\bf h}}
     {t_{N_0}^2 + \sqrt{t_{N_0}^4 + \left( 2 t_0  {\bf a}_0^T {\bf h} \right)^2}} \, .
\label{eq:xi0}
\end{equation}

Therefore, by the construction described above, the deviation of the midpoint satisfies, 
\begin{equation}
\xi_0({\bf h}) = \xi_{\text{ZO}}({\bf h})\, ,
\end{equation}
whenever equation~(\ref{eq:azo_ref}) holds. This condition guarantees that the CRP kinematic geometry remains consistent with the virtual-ZO configuration used as the physical reference.

In summary, we define the midpoint displacement in the CRP domain, $\Delta{\bf m}_{\text{CRP}}$, as
\begin{equation}
\Delta{\bf m}_{CRP}({\bf h}; t_0, {\bf h}_0, {\bf a}_0, V_0) = \xi_0({\bf h}){\bf h}\, ,
\end{equation}
which provides a parametric description of how the midpoint varies with offset and with the underlying reference and kinematic parameters $(t_0, {\bf h}_0, {\bf a}_0, V_0)$.

Using this definition, the following identities follow directly:
\begin{equation}
\begin{array}{lcl}
{\bf m}_0 - {\bf m}_\text{ZO} &=& \Delta{\bf m}_\text{CRP}({\bf h}_0; t_0,{\bf h}_0,{\bf a}_0,V_0)\, , \\
{\bf m}_{CRP}({\bf h}) - {\bf m}_\text{ZO} &=& \Delta{\bf m}_\text{CRP}({\bf h}; t_0,{\bf h}_0,{\bf a}_0,V_0)\, .
\end{array}
\end{equation}

Since ${\bf m}_0 = {\bf m}_{\text{CRP}}({\bf h}_0)$, the midpoint CRP trajectory is expressed explicitly as
\begin{equation}
{\bf m}_\text{CRP}({\bf h}) = {\bf m}_0 + \xi_0({\bf h}){\bf h} - \xi_0({\bf h}_0){\bf h}_0\, .
\label{gCRP}
\end{equation}
This expression corresponds to the mapping from the midpoint reference configuration to the CRP midpoint domain. In the context of seismic mapping, it represents the geometric transformation associated with migration and demigration between two different kinds of finite-offset configurations, as discussed in standard references on midpoint transforms and NMO-like mappings.

In order to make the link between the squared traveltime $t_N^2$ at the half-offset vector ${\bf h}$ with the reference value $t_{N_0}^2$ at ${\bf h}_0$, based on equation~(\ref{eq:tzodmo}), we set the associated NMO-like relation along the CRP trajectory as
\begin{equation}
t_N^2 = \left(\frac{1 - [\xi_0({\bf h}_0)]^2}{1 - [\xi_0({\bf h})]^2}\right)t_{N_0}^2\, .
\label{eq:DMO_CO}
\end{equation}

Therefore, in equation~(\ref{eq:DMO_CO}), the ratio between the two NMO-like quantities acts as a DMO-type operator evaluated at the initial and final half-offsets. This ratio effectively maps the kinematic behavior of the reflection event from one offset configuration to another, correcting for the dip-dependent traveltime. In other words, it serves as a local DMO transformation, regardless of whether the medium is heterogeneous or anisotropic, that links the departure and arrival offsets along the CRP trajectory, ensuring the event remains dynamically consistent under offset continuation.

Furthermore, taking into account the equations~(\ref{eq:DMO_CO}) and~(\ref{eq:tN2}), which is used with respect to the available data in non-zero offsets, we obtain the following expression, which provides an approximation to the CRP traveltime surface as
\begin{equation}
t_\text{CRP}({\bf h}) = \sqrt{t_N^2({\bf h}) + {\bf h}^T{\bf S}_0{\bf h}}\, .
\label{eq:tCRP}
\end{equation}
This formulation shows that the CRP traveltime can be interpreted as a combination of an NMO-like term, $t_N({\bf h})$, and a correction associated with the local wavefront curvature encoded in the quadratic form ${\bf h}^T{\bf S}_0{\bf h}$.

Furthermore, when the conditions expressed in equation~(\ref{eq:xaster}) are satisfied, together with the requirement ${\bf S}_0={\bf S}_*$, we obtain the follow identity
\begin{equation}
t_D({\bf m}_\text{CRP}({\bf h}),{\bf h};{\bf x}_*) \equiv t_\text{CRP}({\bf h})\, ,
\end{equation}
which establishes that the CRP-based approximation is locally consistent with the DSR traveltime operator evaluated at the mapped midpoint ${\bf m}_\text{CRP}({\bf h})$ from $({\bf m}_0,{\bf h}_0)$. This result confirms that, under these approximation conditions, the CRP traveltime provides a physically coherent representation of the underlying reflection geometry, offering a kinematically accurate approximation suitable for stacking, interpolation, and multiparameter analysis. Furthermore, it complies with the definition imposed by the conditions in equation~(\ref{eq:CRP_def}).

To introduce the final slope–continuation relation within the CRP framework, we apply equation~(\ref{eq:azo_ref}) to two distinct half-offsets: the reference and an arbitrary. By equality ${\bf a}_\text{ZO}/t_\text{ZO}$ in both resulting expressions and exploiting the invariance of the underlying kinematic structure along the CRP trajectory, we obtain an explicit mapping that relates the slope at the reference configuration to that at an arbitrary half-offset. Accordingly, we write the slope-continuation operator as
\begin{equation}
{\bf a}_\text{CRP} = \frac{t_{N}^2}{t_{N_0}^2}\frac{t_0}{t_\text{CRP}}{\bf a}_0\, .
\label{eq:slope_cont}
\end{equation}
This relation expresses how the local slope of the traveltime surface evolves as the midpoint migrates along the CRP mapping path.

\begin{figure}
    \centering
    \includegraphics[width=0.8\linewidth]{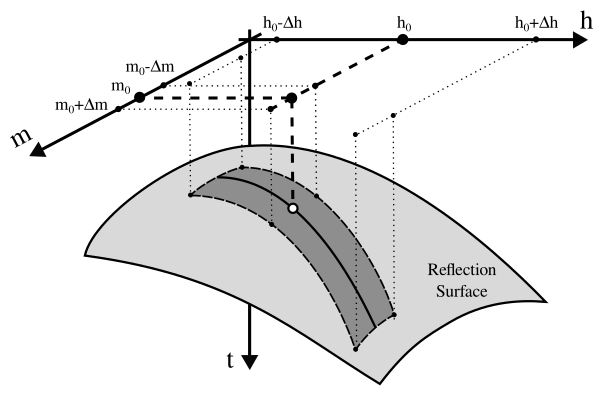}
    \caption{Midpoint and half-offset apertures $(2\Delta m, 2\Delta h)$ in the time domain for a sample located at $(m_0, h_0, t_0)$ over a reflection surface for a 2D acquisition. Note that the OCT trajectory varies across midpoints as a function of half-offset.
}
    \label{fig:crp2d-time}
\end{figure}

Taking all continuation operators for the CRP (see equations~(\ref{gCRP}), (\ref{eq:tCRP}), and~(\ref{eq:slope_cont})), we can write the OCT hypersurface, 5D traveltime operator, as a function of $\Delta{\bf m}$ and ${\bf h}$ as
\begin{equation}
{\cal T}(\Delta{\bf m},{\bf h};t_0,{\bf m}_0,{\bf h}_0) = (t_\text{OCT}(\Delta{\bf m},{\bf h}),{\bf m}_\text{OCT}(\Delta{\bf m},{\bf h}),{\bf h})\, , \end{equation}
where
\begin{eqnarray}
{\bf m}_\text{OCT}(\Delta{\bf m},{\bf h}) &=& {\bf m}_\text{CRP}({\bf h}) + \Delta{\bf m}\, ,\\
t_\text{OCT}(\Delta{\bf m},{\bf h}) &=&  t_\text{CRP}({\bf h}) + 
{\bf a}_\text{CRP}({\bf h})^T\Delta{\bf m}\, .
\label{tCRP4}
\end{eqnarray}
Figure \ref{fig:crp2d-time} illustrates a schematic representation of the OCT trajectory in a 3D dataset, highlighting its underlying geometric structure.

Finally, to reduce the number of degrees of freedom in the operator, we approximate the quadratic form ${\bf h}^T{\bf S}_0{\bf h}$ by $4\|{\bf h}\|^2/V_0^2$, effectively decreasing the parameter count from three to one. Under this approximation, we introduce the auxiliary traveltime function
\begin{equation}
t_\phi({\bf h}) = \sqrt{t_N^2({\bf h}) + \frac{4\|{\bf h}\|^2}{V_0^2}}\, ,
\label{tCRP2}
\end{equation}
which allows us to replace $t_\text{CRP}$ with the simplified expression $t_\phi$. Consequently, we can write the relationship between the exact CRP traveltime and its reduced-parameter approximation as
\begin{equation}
t_\text{CRP}({\bf h})^2 = t_\phi({\bf h})^2 + {\bf h}^T\left({\bf S}_0 - \frac{4{\bf I}}{V_0^2}\right){\bf h}\, ,
\end{equation}
highlighting that the difference between the two quantities is governed entirely by the residual curvature term modulated by the matrix ${\bf S}_0 - 4{\bf I}/V_0^2$. This simplification is particularly advantageous because it reduces the computational cost associated with estimating the kinematic parameters by relying on the traveltime $t_\phi({\bf h})$. However, the deviation has a quadratic term in the offsets. In summary, such a deviation happens when the eigenvalues of the matrix ${\bf S}_0$ are different due to its matrix structure, and ${\bf h}$ is not parallel to ${\bf h}_0$. On the other hand, when the eigenvalues are equal, the deviation is null regardless of ${\bf h}$ values.

\section{Experiments and numerical results}

To emulate and evaluate scenarios of this kind, we use both finite-difference modeled data and a real onshore seismic dataset affected by acquisition restrictions. For the synthetic dataset, the experiment focuses exclusively on interpolating the data onto a new, entirely regular acquisition grid. For the real dataset, we apply the proposed method to perform both interpolation and data enhancement, addressing issues such as incomplete spatial sampling, irregular coverage, and amplitude inconsistencies.

The experiments were conducted on the HPG Lab cluster at UNICAMP. The computing infrastructure comprises 18 Intel Xeon Gold 6148 CPU cores and 22 NVIDIA V100 GPUs, each with 16~GB of memory. Dataset manipulation, preprocessing, and visualization were performed using Shearwater Reveal.

\subsection{Scheme for interpolation, regularization, and enhancement}

We now describe the proposed scheme for interpolating, regularizing, and enhancing prestack seismic data. In this formulation, we assume an input prestack dataset with an arbitrary acquisition geometry and known recorded amplitudes, and an output prestack dataset defined on a user-specified acquisition geometry, for which no amplitudes have yet been assigned. The task of the OCT-based operator is therefore to compute the amplitudes at the output grid locations such that the resulting dataset simulates what would have been recorded under the new acquisition geometry, including the amplitude-enhancement effect introduced by our method. To accomplish this, we first provide a concise description of the parameter-search step.

As previously mentioned, let the input and output prestack datasets be defined on two distinct grids. For each point on the output grid, the corresponding set of OCT parameters is estimated by analyzing the input dataset. These estimations are performed through a multiparameter similarity analysis in the domain of user-selected kinematic parameters, using the approach known as Evolution by Neighborhood Similarity (ENS), as presented by \cite{Okita:2024}. In this work, for the ENS setup, we adopt a 2-4-2 forward-backward-forward configuration and an ENS grid of size $16 \times 16$ with 16 individuals per interactive domain, which totals 128 coherence computations per sample. For each candidate element in this parameter set, the similarity is computed along the OCT traveltime, with the reference point being the output-grid location under analysis. The similarity calculation uses all available half-offsets for which the OCT operator gathers valid traces. A global optimization algorithm is then applied to identify the parameter vector that maximizes the similarity measure. In the case of the OCT operator, the relevant parameter set is $\{a_x, a_y, V_0\}$, where $(a_x, a_y)={\bf a}_0$. The output amplitude assigned to the reference point on the output grid corresponds to the stacking value obtained using this optimal parameter set.

The parameter-estimation strategy summarized above is widely used in several stacking-based seismic processing and imaging techniques and is well documented in the literature, for example, in \cite{Ribeiro:2020} and \cite{Ribeiro:2023:ENS}. It is important to note that a detailed discussion of these estimation methods lies beyond the scope of this paper.

Once the OCT parameters have been estimated at all output-grid locations, we can describe the proposed OCT-based interpolation scheme. The method is designed to accomplish two main goals: (i) enhance the continuity and coherence of seismic events, and  (ii) reconstruct data on a regularized grid consistent with a user-defined acquisition geometry.

The amplitude assigned to each output-grid location is computed in two stages, following the strategy of \cite{Coimbra:2015:stack} and \cite{coimbra:2019enhancement}. In the first stage, the output amplitude results from stacking the input data along an appropriate aperture in the midpoint and offset directions in the neighborhood of the OCT traveltime corresponding to the reference point, see \cite{Faccipieri:2016} for an analysis of appropriate apertures. Because this stage uses the actual amplitudes from the input dataset, a sufficiently small and properly chosen aperture ensures that the stacked amplitude closely approximates the amplitude that would be observed if the event had been recorded under the output acquisition geometry. In the second stage, a spreading or refinement step is applied to further improve continuity and stabilize amplitudes. Following the approach of \cite{coimbra:2019enhancement}, the output volume is obtained by spreading the stacked energy from each input sample over a predefined aperture and weighting it according to a physically motivated kernel. This weighting ensures that the resulting amplitudes represent compatible averages of the contributing input amplitudes. Constructive interference reinforces coherent seismic events, thereby enhancing their amplitude, while residual incoherent noise is attenuated through destructive and constructive interferences.

\subsection{Synthetic data example}

To assess the performance of the proposed OCT-based regularization and interpolation strategy under controlled conditions, we apply the method to a 5D synthetic seismic dataset generated by finite-difference modeling. The synthetic experiment is designed to reproduce a land-acquisition scenario while maintaining complete control over the subsurface velocity model and acquisition parameters.

\begin{figure}
    \centering
    \includegraphics[width=0.95\linewidth]{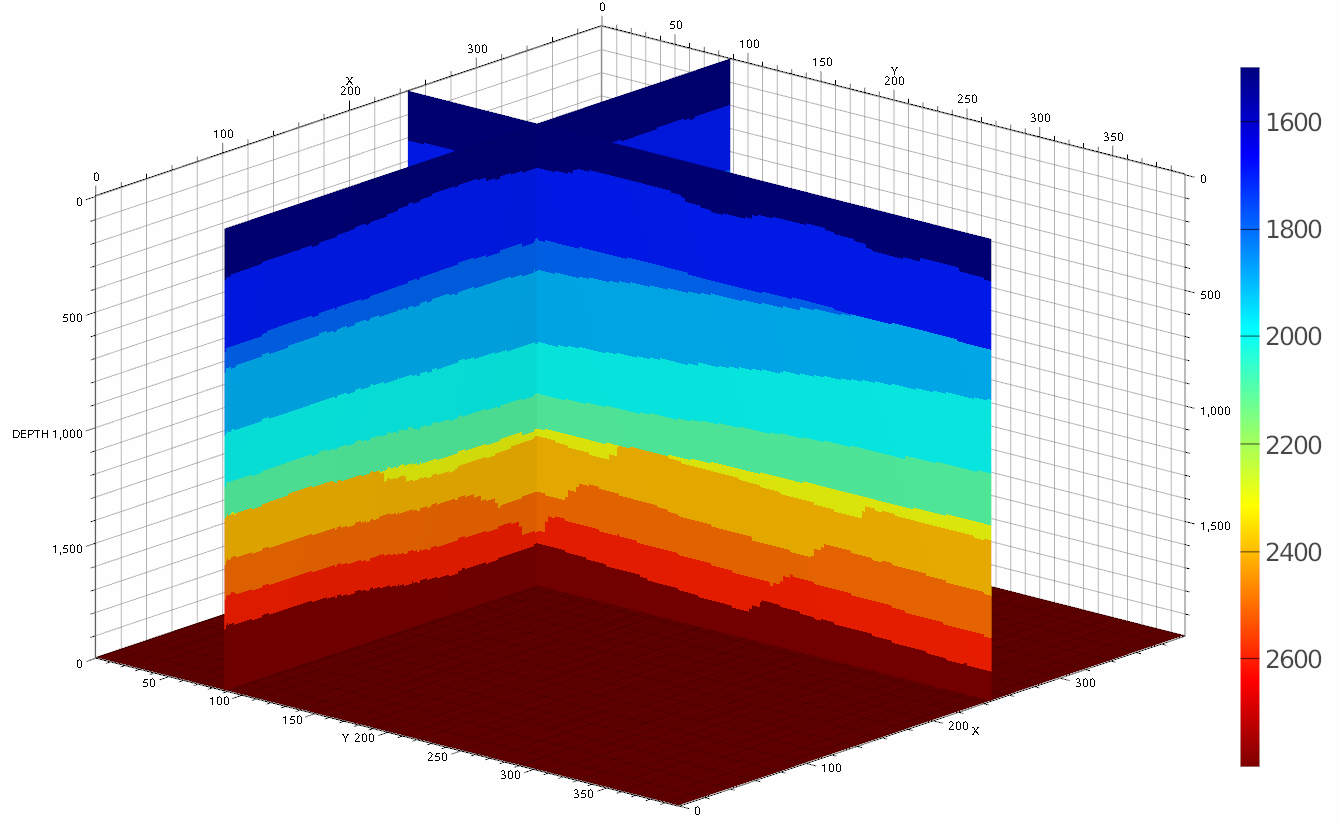}
    \caption{Synthetic velocity model in depth (m/s) used for dataset simulation and algorithm validation.}
    \label{fig:ex1-depth}
\end{figure}

The subsurface model consists of an isotropic laterally heterogeneous depth-velocity distribution, as illustrated in Figure~\ref{fig:ex1-depth}. The velocity model is defined on a regular Cartesian grid with a spatial discretization of $8\,\mathrm{m}$ in the $x$, $y$, and $z$ directions. The horizontal extent of the model comprises 400 by 400 grid points, corresponding to an area of $3200\,\mathrm{m}$ by $3200\,\mathrm{m}$, while the vertical dimension includes 250 grid points, reaching a maximum depth of $2000\,\mathrm{m}$. Velocity values vary between $1500\,\mathrm{m/s}$ and $2800\,\mathrm{m/s}$, producing a moderately complex propagation medium with both vertical velocity gradients and lateral velocity variations that are representative of typical land seismic environments.

\begin{figure}
    \centering
    \includegraphics[width=1\linewidth]{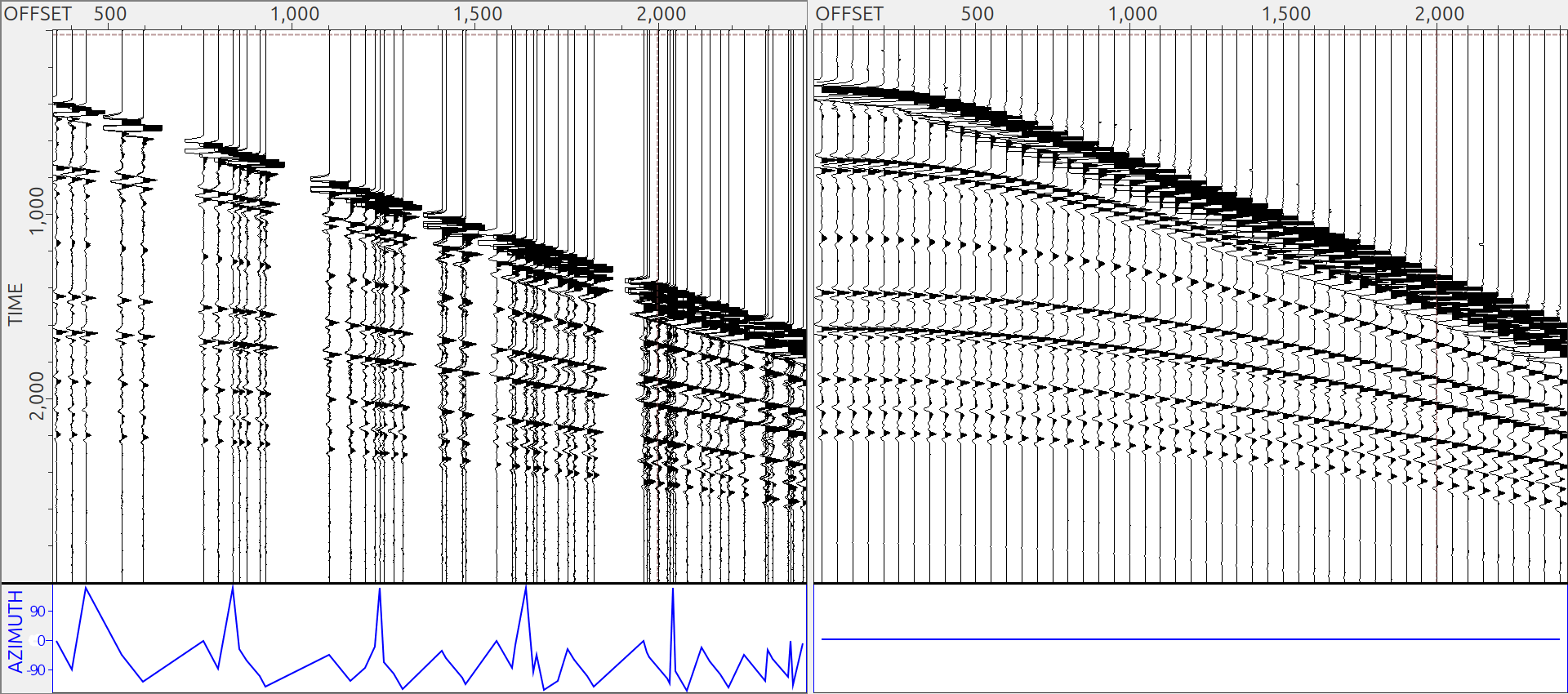}
    \caption{Comparison of CMP gathers from the original dataset (left) and the OCT regularization (right). The original data exhibits significant gaps and non-uniformity in offset and azimuth distribution, which are effectively mitigated by the OCT regularization.}
    \label{fig:ex1-cmps}
\end{figure}

Synthetic seismic data are generated using an acoustic finite-difference time-domain scheme, ensuring accurate simulation of wave-propagation effects, including reflection moveout, diffraction responses, and amplitude variations associated with velocity heterogeneities. A Ricker wavelet with a dominant frequency of $15\,\mathrm{Hz}$ is used as the source signature, providing sufficient bandwidth to resolve reflector continuity and diffraction events while remaining consistent with field-scale land data.

\begin{figure}
    \centering
    \includegraphics[width=1\linewidth]{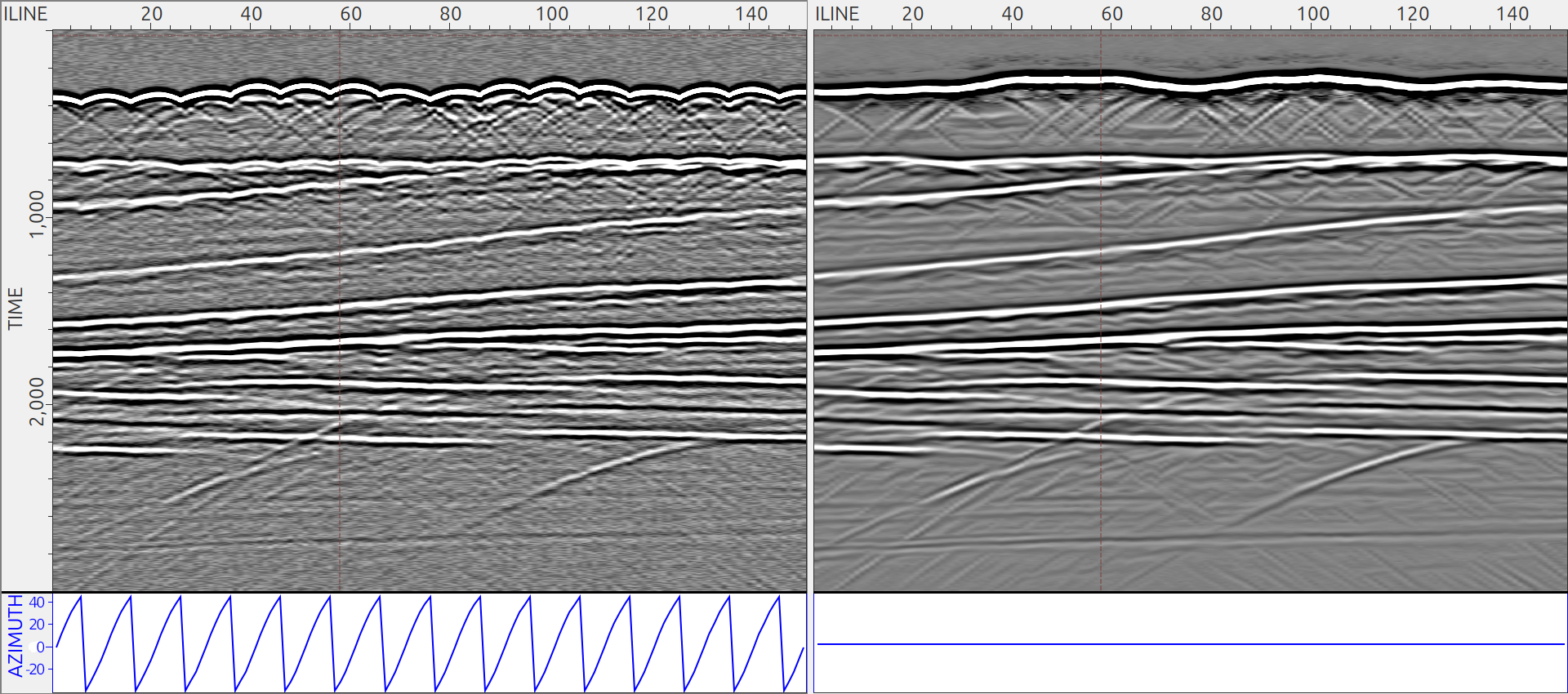}
    \caption{Inline comparison between the original dataset with multi-azimuth and an offset bin ranging from 100 to 400 m (left) and the OCT-regularized ZO and zero azimuth (right).}
    \label{fig:ex1-il-zo}
\end{figure}

The acquisition geometry follows a land configuration with inlines aligned with azimuth zero, corresponding to a conventional orthogonal layout. We deployed the sources on a regular grid with spacing of $200\,\mathrm{m} \times 200\,\mathrm{m}$, resulting in a total of $16 \times 16$ source locations distributed across the model surface. Receivers are placed with a denser spacing of $40\,\mathrm{m} \times 40\,\mathrm{m}$, yielding $76 \times 76$ receiver positions. This configuration produces a broad range of offsets and midpoint locations, allowing the evaluation of the proposed method in a five-dimensional prestack domain.

\begin{figure}
    \centering
    \includegraphics[width=1\linewidth]{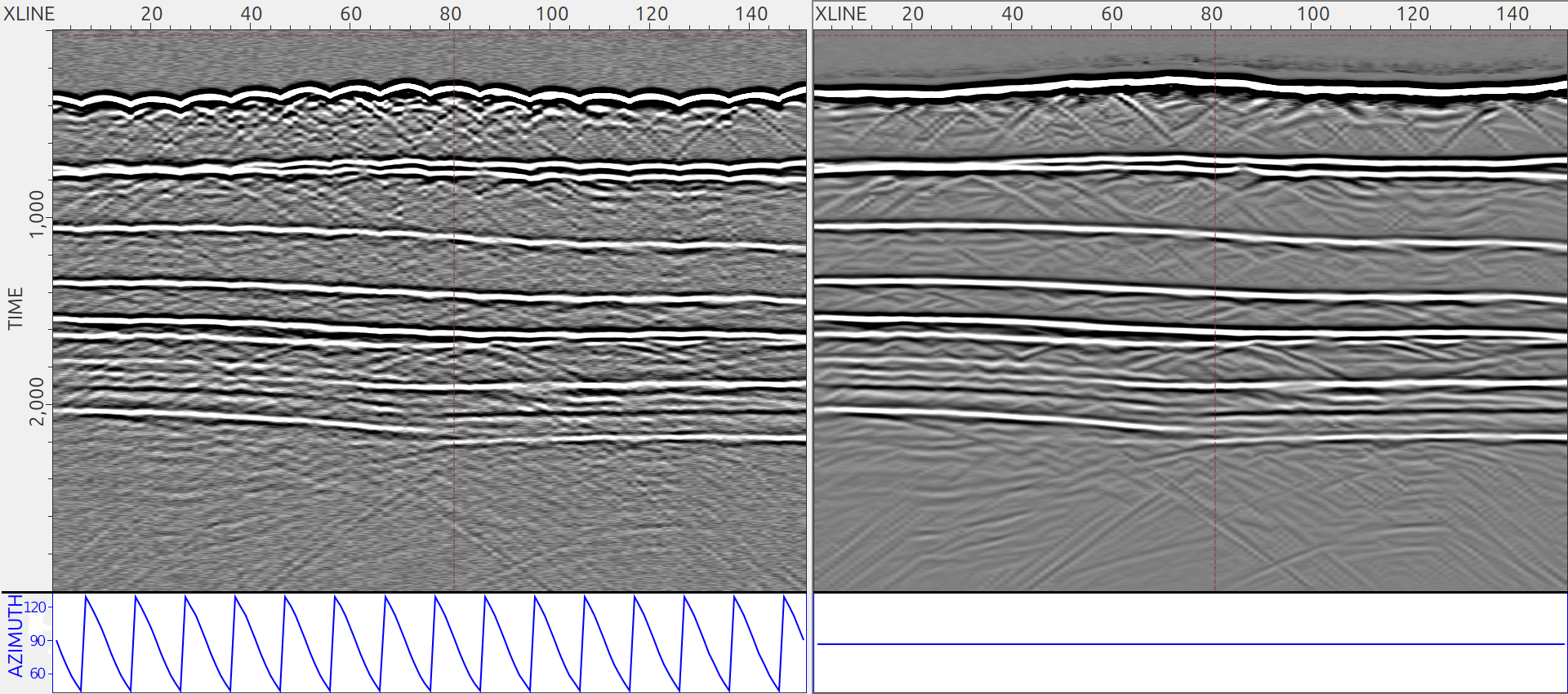}
    \caption{Crossline comparison between the original dataset with multi-azimuth and an offset bin ranging from 100 to 400 m (left) and the OCT-regularized ZO and azimuth 90° (right).}
    \label{fig:ex1-xl-zo}
\end{figure}

The resulting prestack dataset is organized in terms of common midpoint (CMP), inline (IL), and crossline (XL) coordinates, with CMP bin sizes of $20\,\mathrm{m}$ by $20\,\mathrm{m}$, covering 151 inline and 151 crossline positions. The combination of relatively sparse source spacing and denser receiver sampling results in a nonuniform fold distribution and incomplete coverage in some areas of the CMP domain, thereby providing a suitable benchmark for evaluating the regularization and reconstruction capabilities of the proposed OCT operator.

\begin{figure}
    \centering
    \includegraphics[width=1\linewidth]{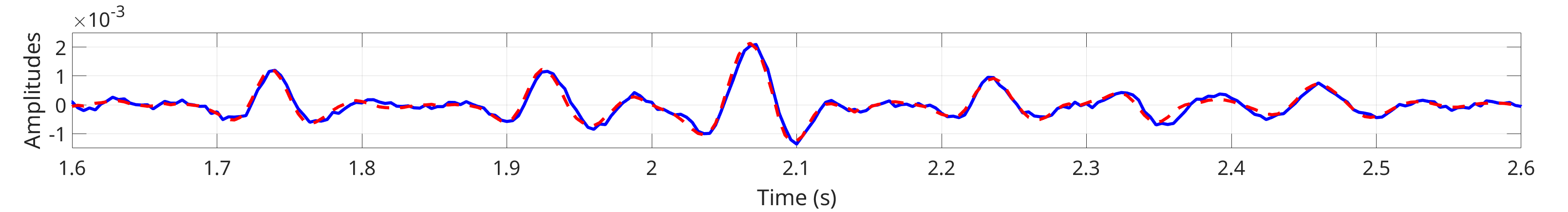}
    \includegraphics[width=1\linewidth]{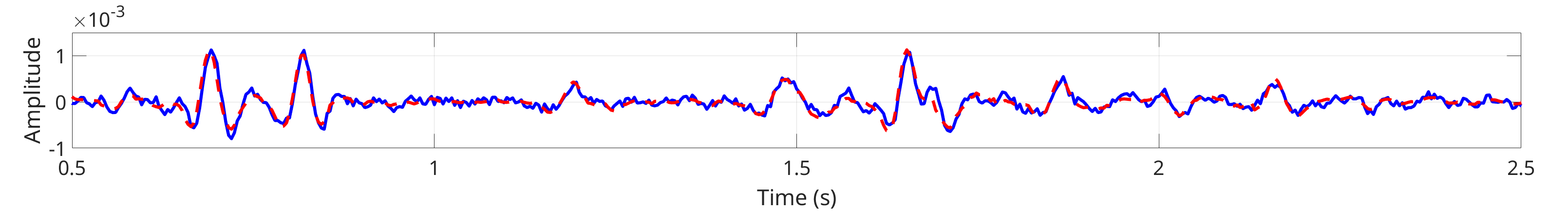}
    \caption{Single trace comparison between the original dataset (solid blue line) and the OCT-regularized (dashed red line). Top: ZO trace at inline 61, crossline 101. Bottom: 2000 m offset trace (zero azimuth) at inline 71, crossline 91.}
    \label{fig:ex1-trc}
\end{figure}

In this synthetic experiment, the primary objective is to interpolate a prestack dataset onto a fully regularized acquisition grid, enhancing event continuity while preserving physically consistent traveltime and amplitude behavior. Because the accurate subsurface model and acquisition geometry are known, the synthetic dataset provides a reliable reference for validating the kinematic accuracy, reconstruction fidelity, and robustness of the proposed method before its application to field data.

\begin{figure}
    \centering
    \includegraphics[width=1\linewidth]{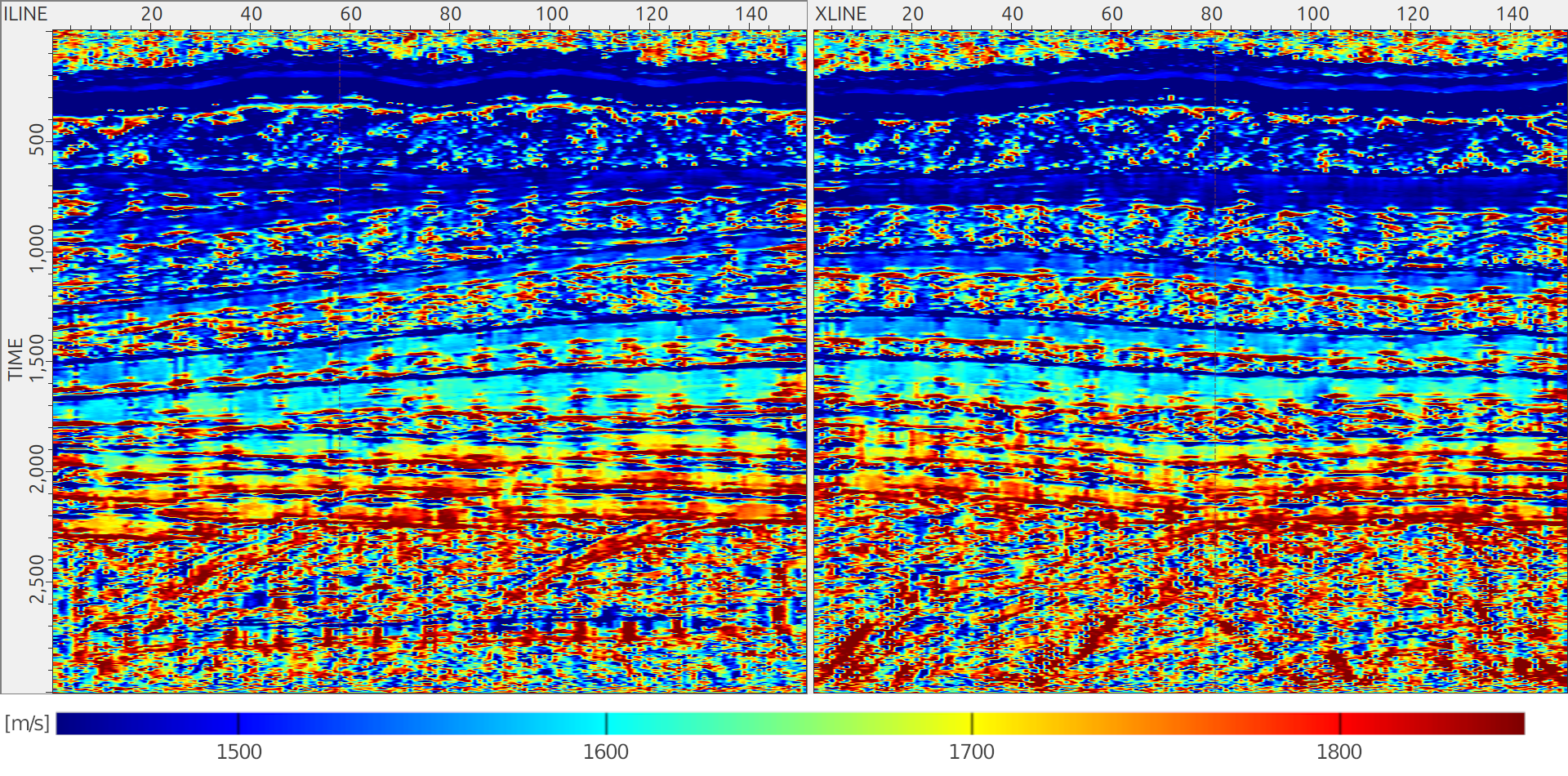}
    \includegraphics[width=1\linewidth]{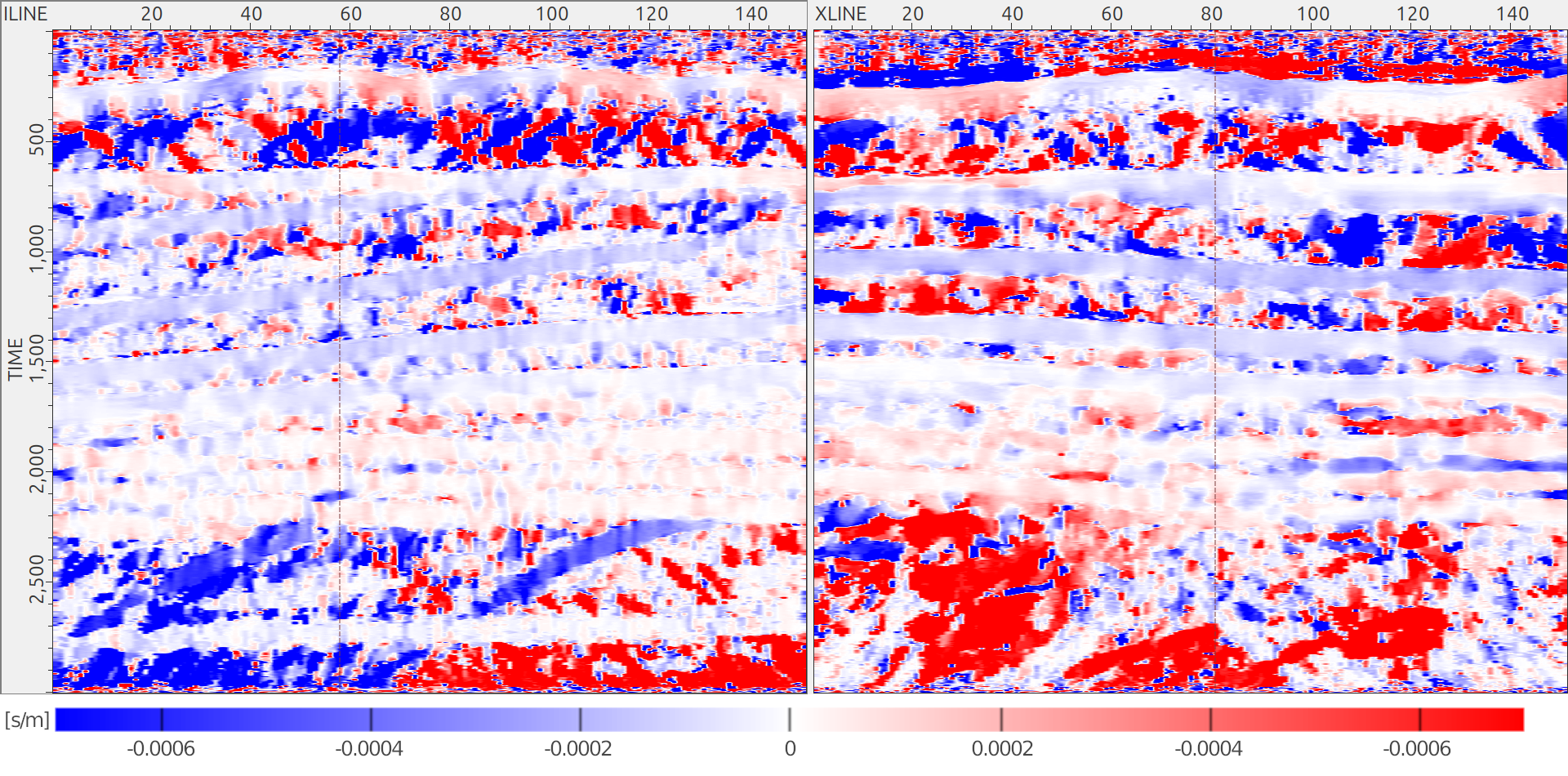}
    \includegraphics[width=1\linewidth]{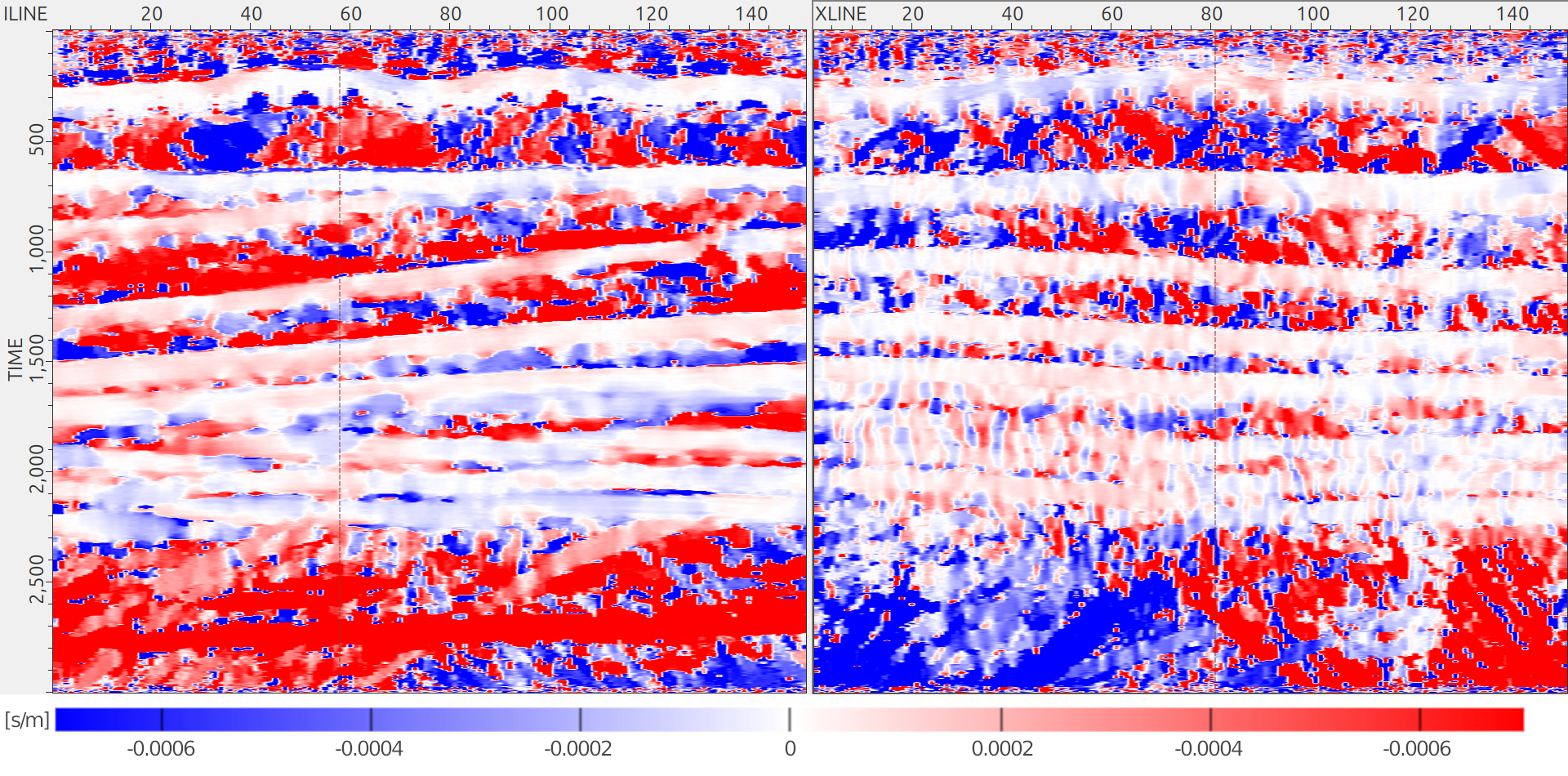}
    \caption{Kinematic parameter $V_0$ ({\color{blue}top}),  $a_x$ ({\color{blue}center}), and $a_y$ ({\color{blue}bottom}) obtained for the ZO section on inline and crossline directions.}
    \label{fig:ex1-params}
\end{figure}

The parameter-estimation and stacking stage uses a midpoint aperture of 25 m and a half-offset aperture of 200 m, centered at each target output location. These apertures define the local neighborhood over which the method evaluates similarity measures and stacks the trajectories. The midpoint aperture controls the lateral coherence of the estimated parameters, whereas the half-offset aperture defines the range of offsets used to compute the OCT kinematics. The chosen values provide sufficient offset diversity to stabilize parameter estimation without introducing kinematic smearing due to excessive spatial averaging.

\begin{figure}[htbp]
    \centering
    \includegraphics[width=1\linewidth]{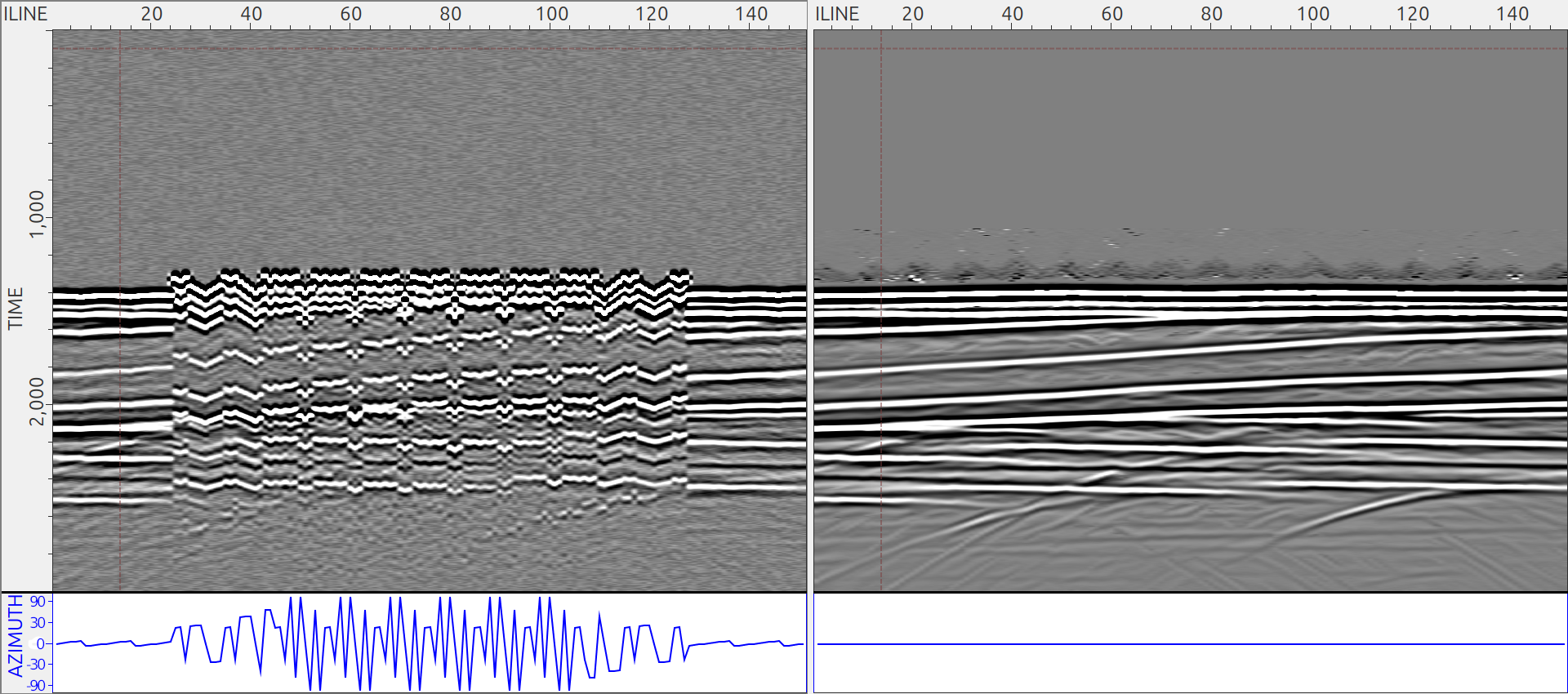}
    \caption{Inline comparison between the original dataset with multi-azimuth and an offset bin ranging from 1800 to 2200 m (left) and the OCT-regularized at offset 2000 m and zero azimuth (right).}
    \label{fig:ex1-il-fo}
\end{figure}

After parameter estimation and stacking, the method applies a spreading operation to enhance event continuity and stabilize amplitudes in the reconstructed prestack volume. The method defines the spreading aperture exclusively in the midpoint domain, with a radius of 15 m. This relatively compact aperture preserves local kinematic accuracy while promoting constructive interference of coherent events and attenuating residual incoherent noise.

\begin{figure}[htbp]
    \centering
    \includegraphics[width=1\linewidth]{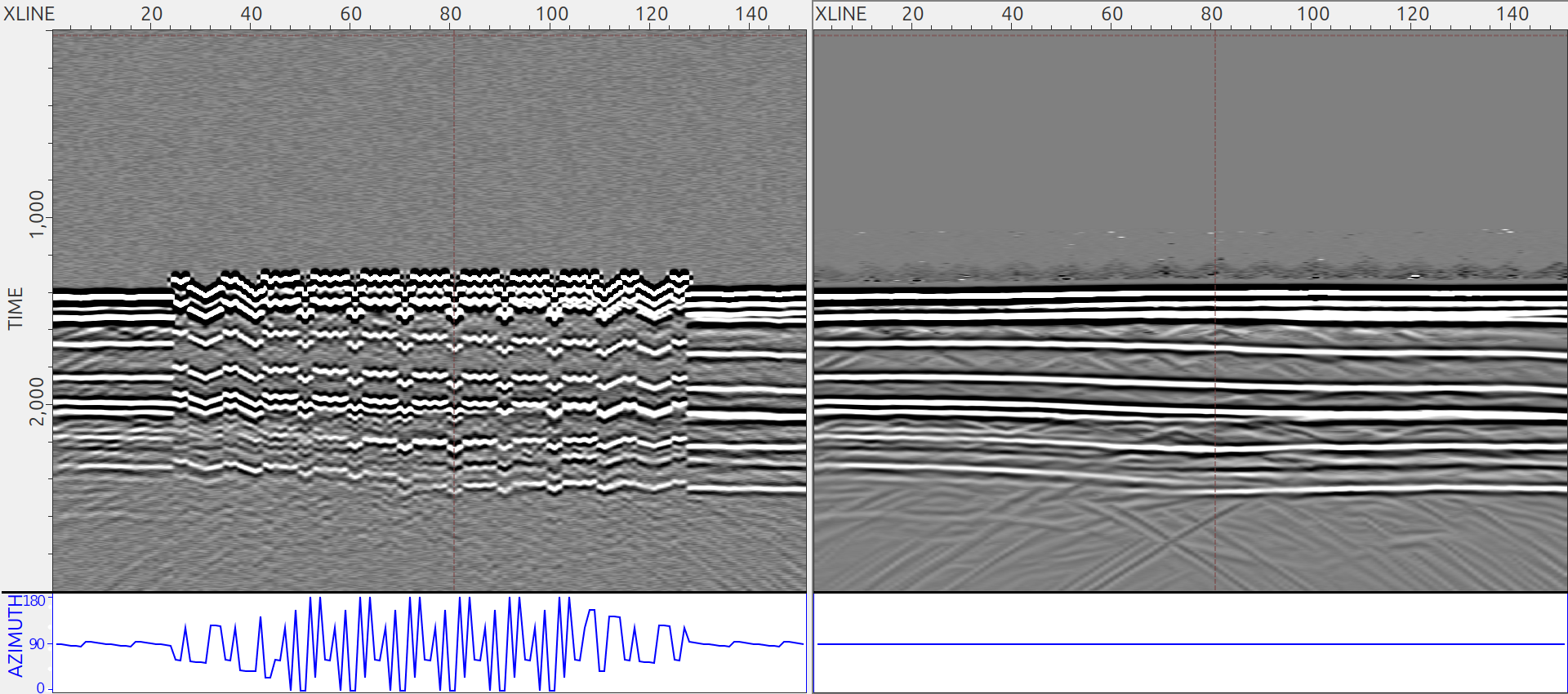}
    \caption{Crossline comparison between the original dataset with multi-azimuth and an offset bin ranging from 1800 to 2200 m (left) and the OCT-regularized at offset 2000 m and azimuth 90° (right).}
    \label{fig:ex1-xl-fo}
\end{figure}

The method generates the output prestack dataset on a regularized acquisition grid defined by IL and XL coordinates, with a spatial sampling of $20\,\mathrm{m}$ by $20\,\mathrm{m}$. This geometry matches the CMP binning used in the synthetic experiment and ensures a uniform midpoint distribution suitable for subsequent processing and imaging. Offsets in the output domain are sampled at regular intervals of 50 m aligned with the zero azimuth, covering a total range from 0 to 2400 m, as shown in Figure~\ref{fig:ex1-cmps}. This offset sampling provides adequate resolution of moveout behavior while maintaining computational tractability in the five-dimensional prestack domain. Together, these parameter choices define a physically consistent and numerically stable configuration for applying the OCT operator in both parameter estimation and data reconstruction. Figure~\ref{fig:ex1-trc} presents a single-trace comparison for two selected offsets, where the reconstructed traces are plotted against their modeled counterparts (ground truth). The first arrival was omitted from the display due to its high amplitude, which would otherwise have obscured the lower-amplitude events and hindered the visual analysis of reconstruction accuracy.

\begin{figure}[htbp]
    \centering
    \includegraphics[width=1\linewidth]{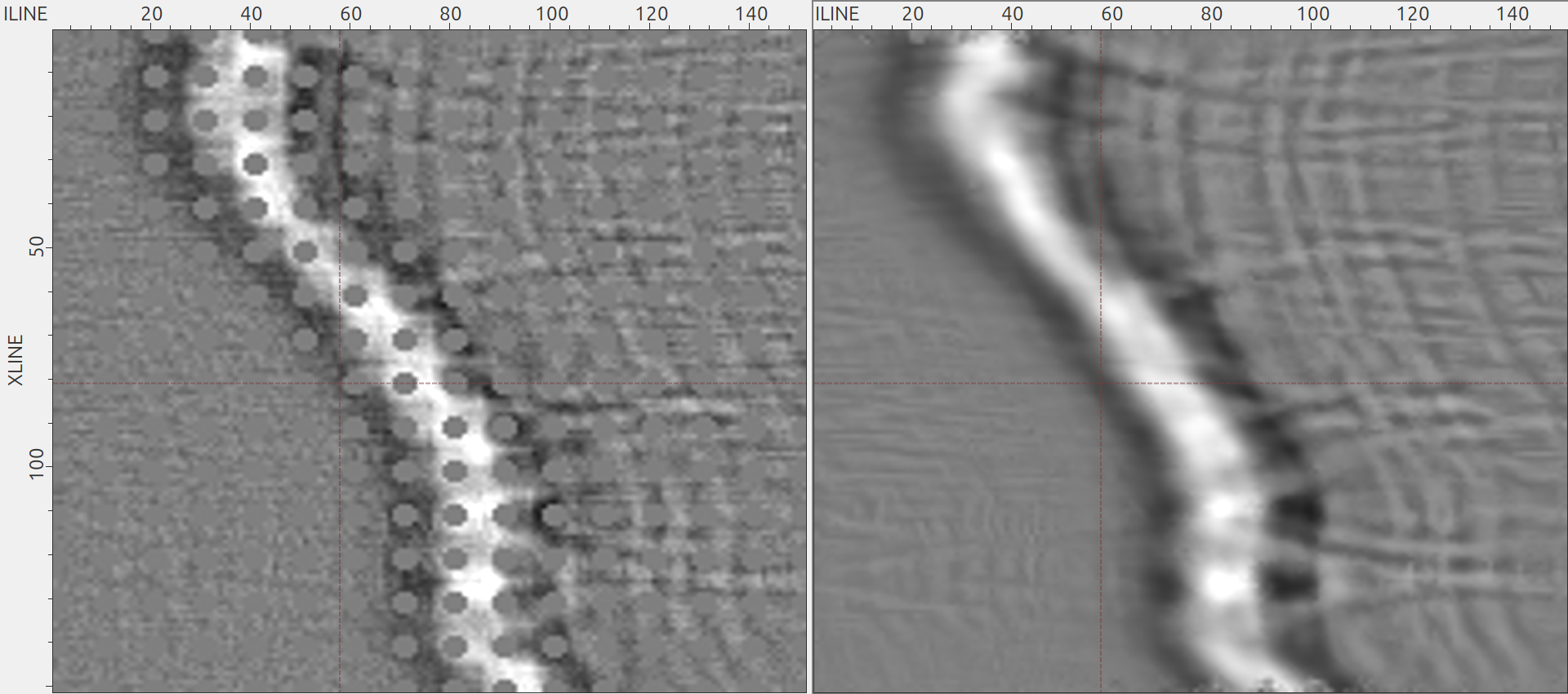}
    \includegraphics[width=1\linewidth]{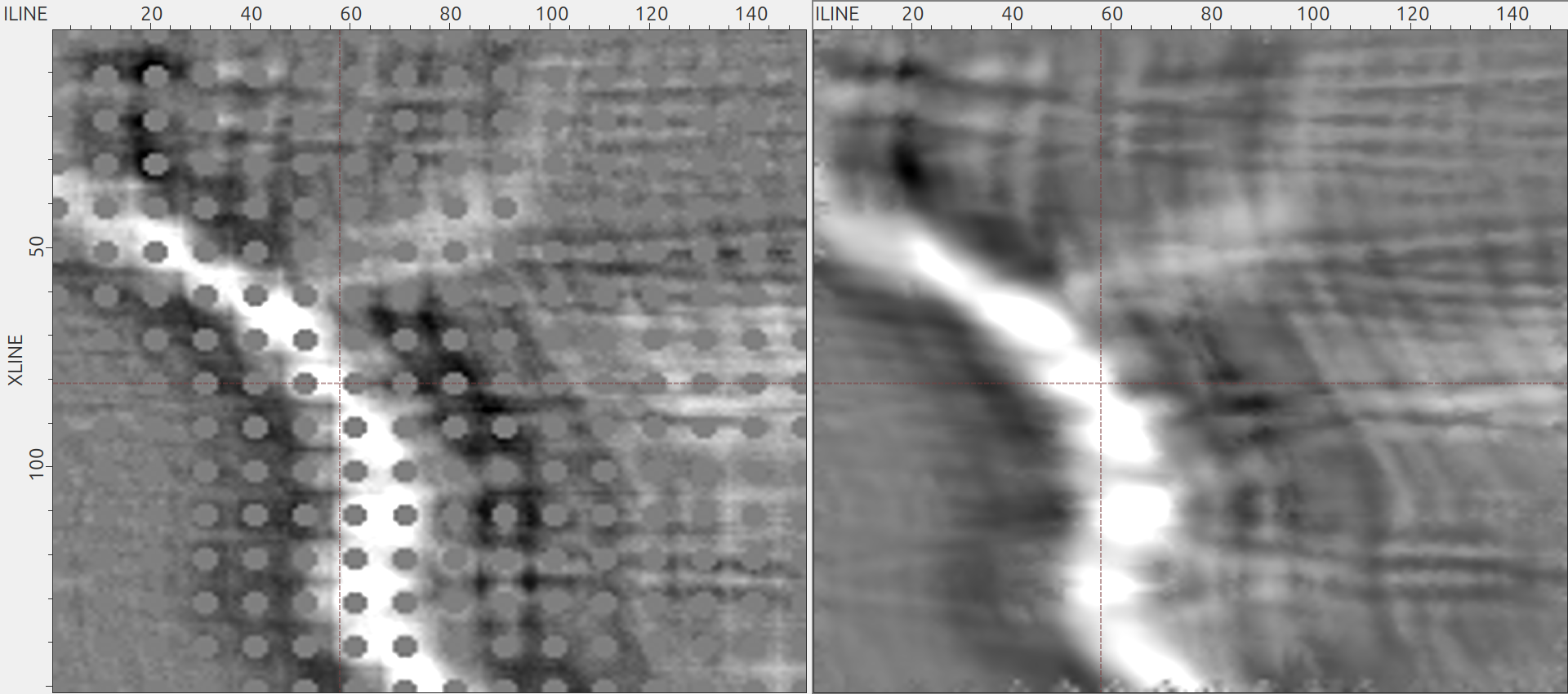}
    \caption{Comparison of time-slices at 1432 (top) and 1652 ms (bottom) from the original dataset with an offset bin ranging from 100 to 400 m (left) and the OCT regularization at ZO section (right).}
    \label{fig:ex1-ts1}
\end{figure}

Figures \ref{fig:ex1-il-zo} and \ref{fig:ex1-xl-zo} show the 100–400 m offset bin for the original inlines and crosslines, compared with their OCT-regularized zero-offset counterparts (reconstructed at zero azimuth). Note that the traveltime shifts associated with the varying offsets and azimuths in the original data were effectively eliminated. The kinematic parameters estimated for the zero-offset case are presented in Figure~\ref{fig:ex1-params}. Consistent results are observed for the 2000 m offset case. Figures \ref{fig:ex1-il-fo} and \ref{fig:ex1-xl-fo} compare the 1800–2200 m offset bin with the regularized results. In this further offset range, traveltime shifts due to offset and azimuth variations are more pronounced. However, the regularization successfully reconstructed the data into the target configuration. Figure~\ref{fig:ex1-ts1} provides a broader perspective on the results, presenting two time slices that compare the original dataset with the OCT-regularized ZO section output. This comparison highlights the spatial continuity and the effectiveness of the regularization across the survey area.

\subsection{Real data example}

\begin{figure}[htbp]
    \centering
    \includegraphics[width=1\linewidth]{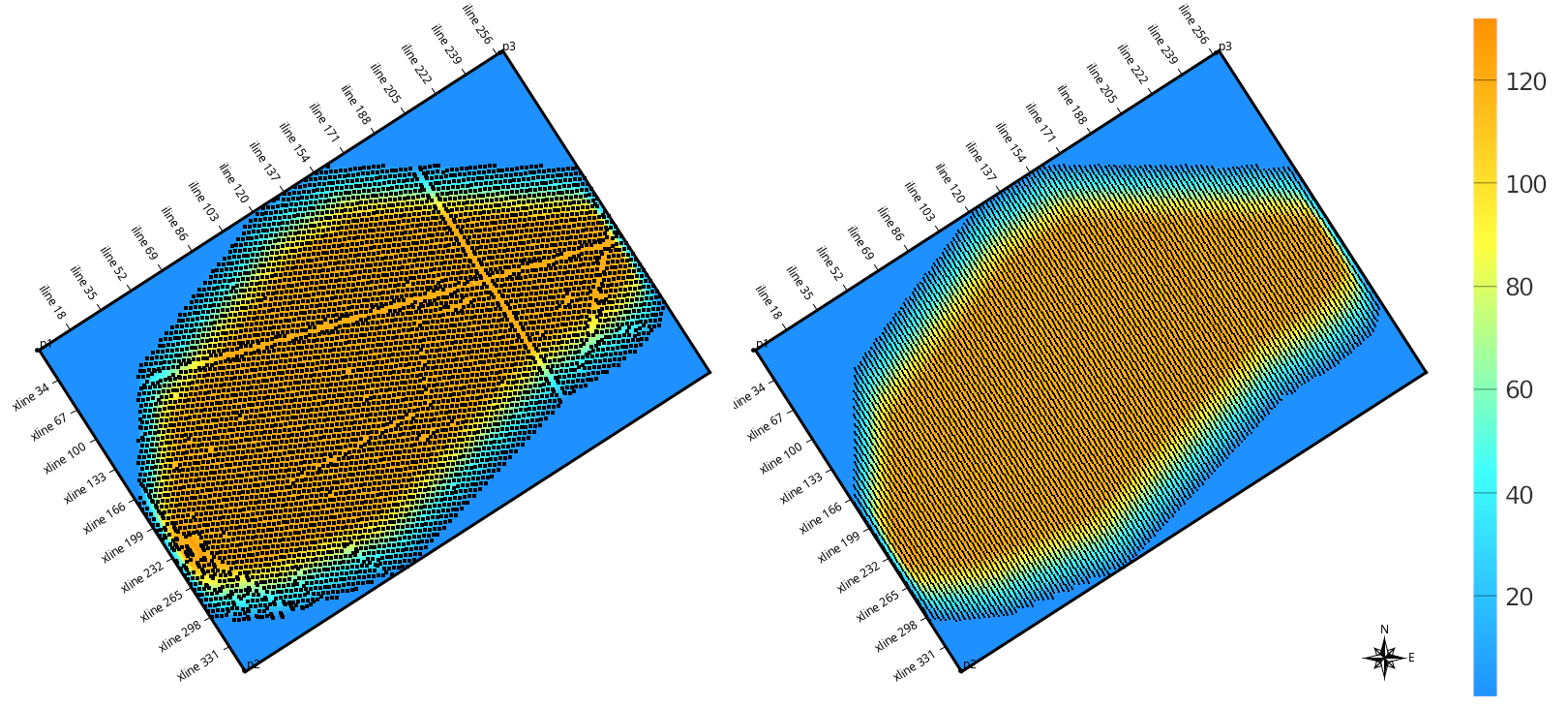}
    \caption{Real dataset acquisition geometry with source (left) and receiver (right) positions, represented by the black squares, together with the fold number for each inline-crossline pair.}
    \label{fig:ex2-geom}
\end{figure}

The proposed method was also tested on a real onshore dataset from the Potiguar Basin, Brazil. As shown in the survey geometry (Figure~\ref{fig:ex2-geom}), the source positions exhibit notable spatial irregularities. The dataset is characterized by an inline spacing of 50 m and a crossline spacing of 25 m.

\begin{figure}[htbp]
    \centering
    \includegraphics[width=1\linewidth]{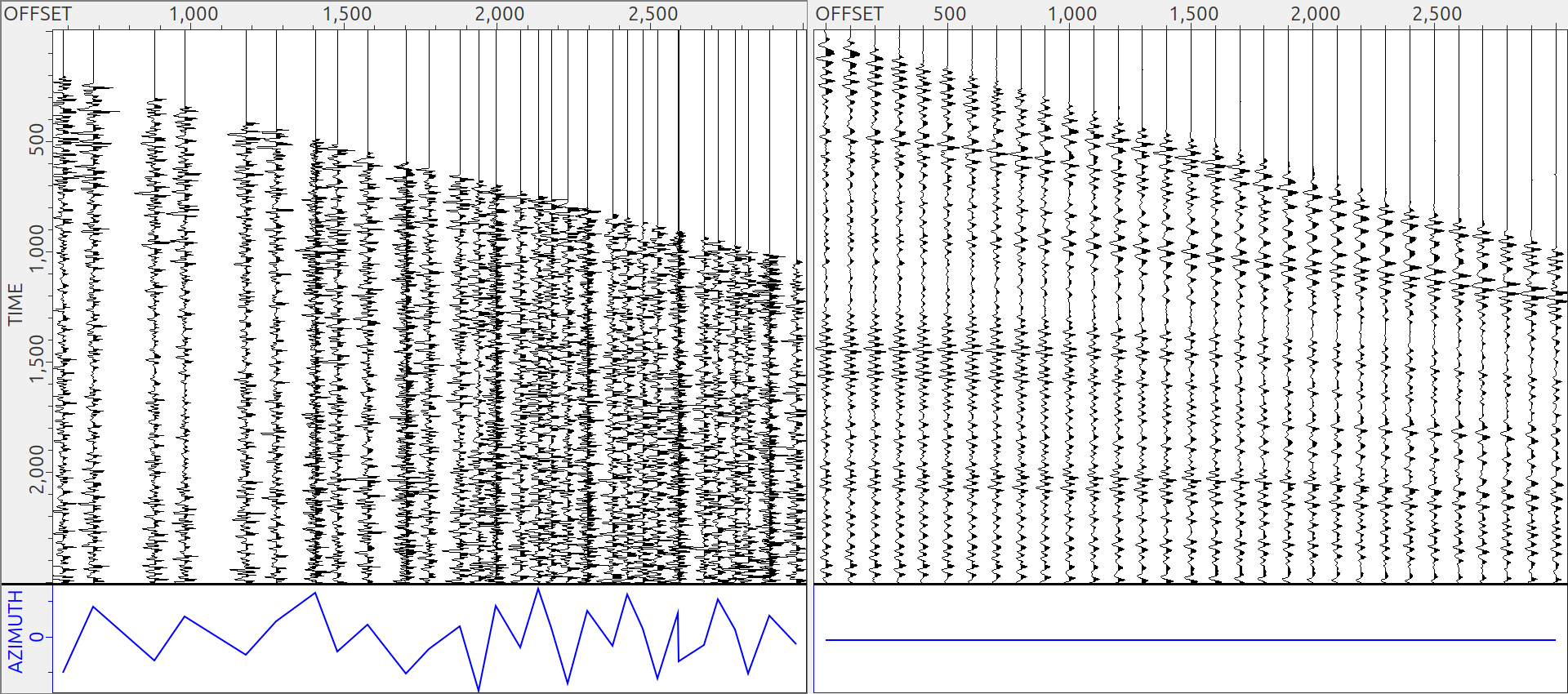}
    \includegraphics[width=1\linewidth]{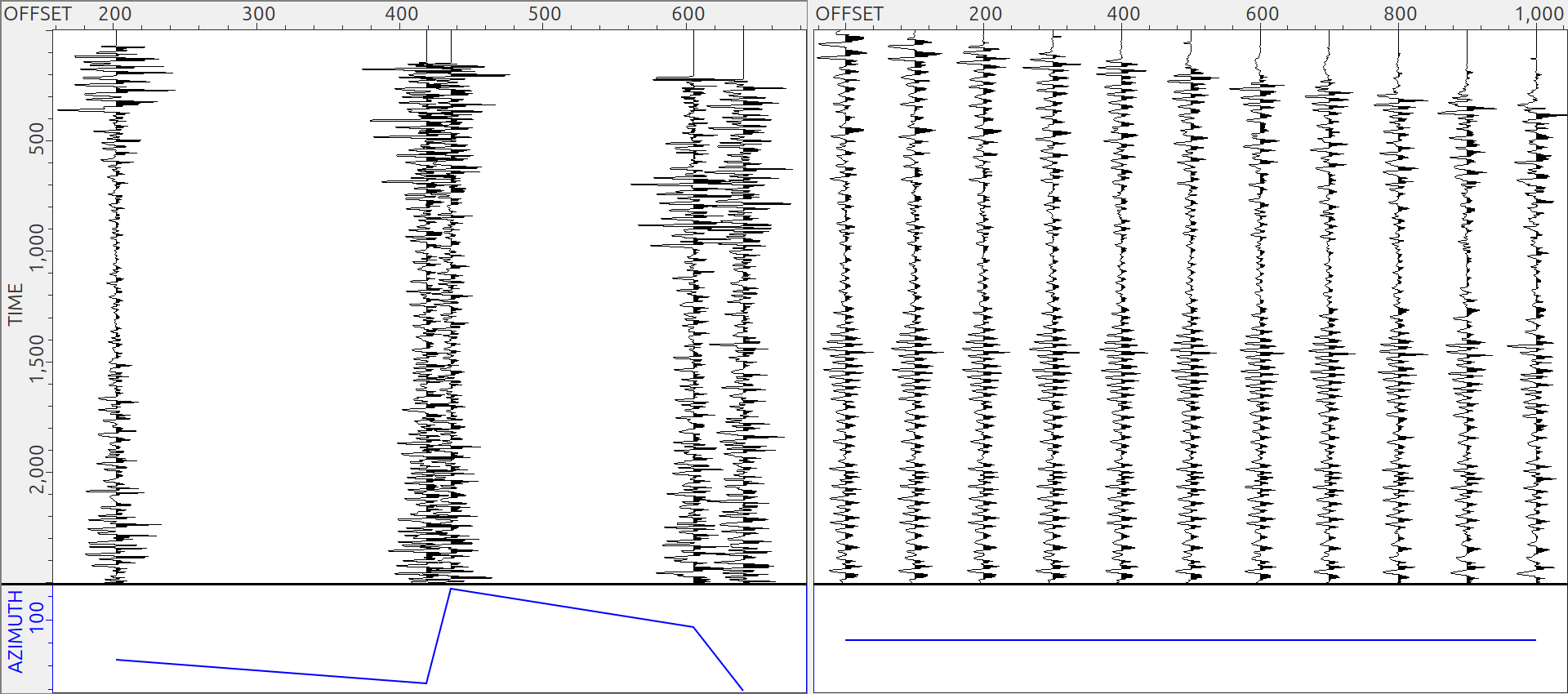}
    \caption{Comparison of CMP gathers from the real dataset (left) and after OCT regularization (right) for traces near the azimuth 0° (top) and 90° (bottom). Consistent with the synthetic results, the real data exhibits significant acquisition footprints, irregular offset-azimuth distribution, and low signal-to-noise ratio. The OCT workflow effectively mitigates these sampling gaps while enhancing reflector continuity and signal coherence.}
    \label{fig:ex2-cmp}
\end{figure}

The OCT-regularization output geometry was designed to maintain the original bin size while regularizing offsets in 100 m increments, ranging from 0 to 3000 m. During the parameter estimation phase, a time-varying aperture of 100–200 m for midpoints and 350–450 m for offsets was employed. The velocity search interval spans from 2500 m/s to 7500 m/s and $\pm$0.0008 s/m for slopes. In order to prevent event smoothing, the stacking apertures were reduced to 50–100 m in the midpoint domain, while the offset apertures were preserved. Figure~\ref{fig:ex2-cmp} compares CMP gathers with traces aligned near zero and 90° azimuths from both the original dataset and its regularized counterpart. Consistent with the synthetic examples, the original data exhibits offset gaps and irregularities that are completely resolved in the regularized output. It should be noted that for this dataset, the varying azimuthal coverage limits the maximum offset range available for reconstruction.

\begin{figure}[htbp]
    \centering
    \includegraphics[width=1\linewidth]{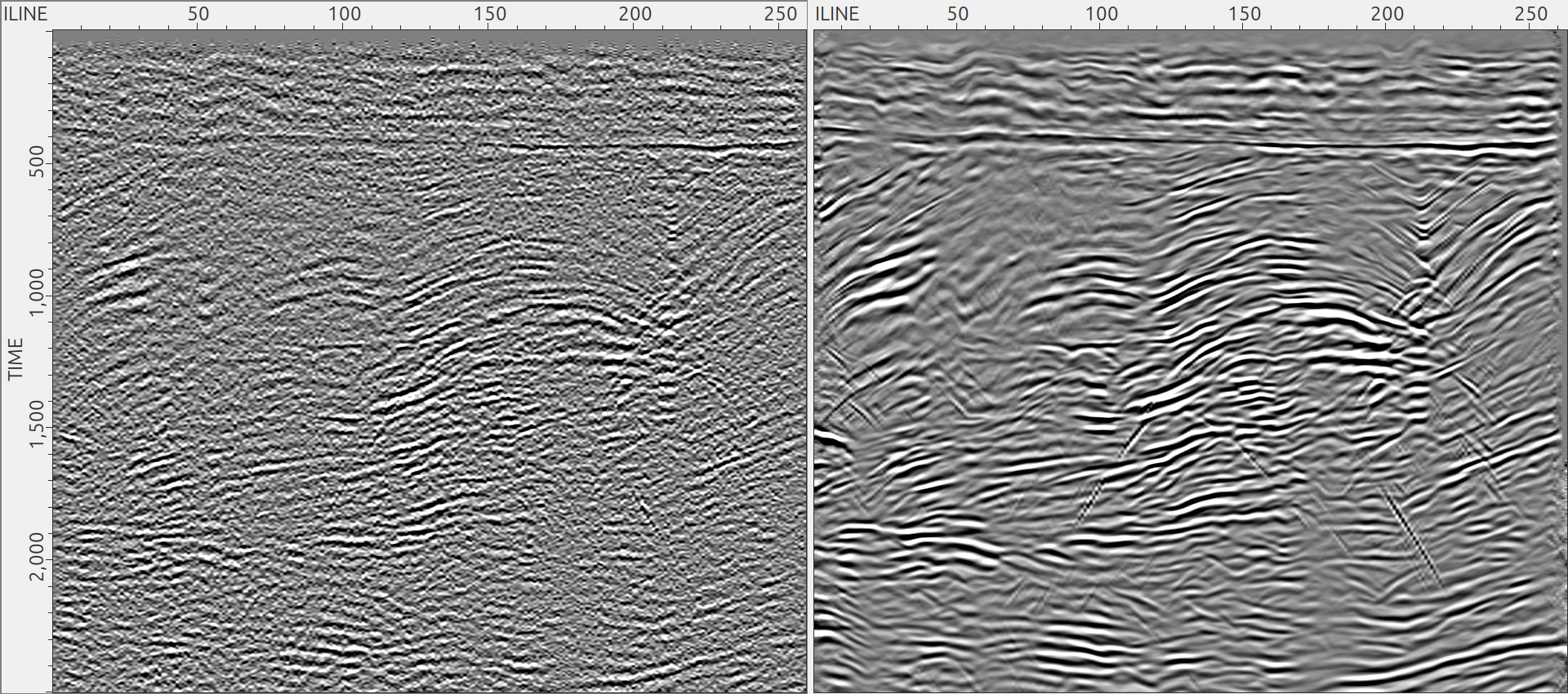}
    \includegraphics[width=1\linewidth]{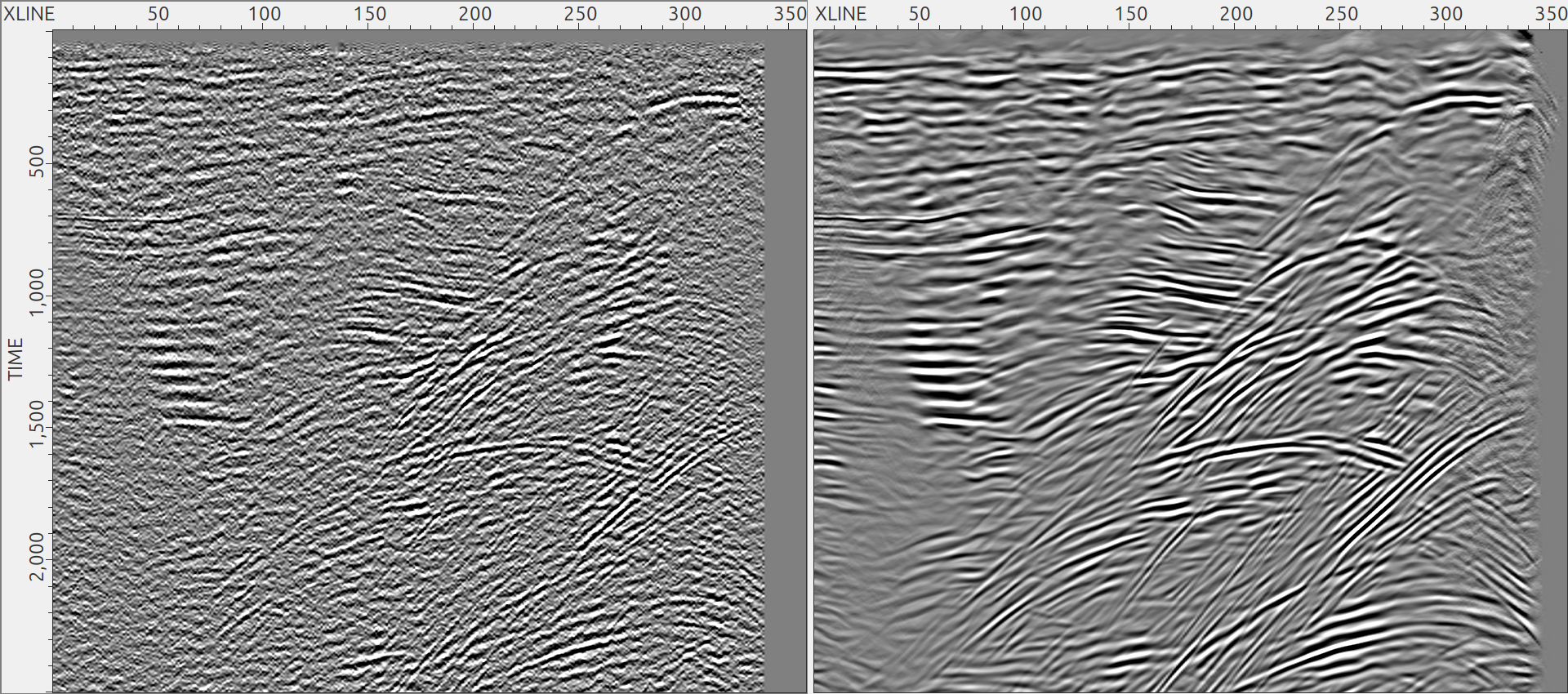}
    \caption{Comparison between stacked sections from the original dataset (left) and the OCT regularization at zero-offset (right) for inline (top) and crossline (bottom) directions.}
    \label{fig:ex2-stks}
\end{figure}

\begin{figure}[htbp]
    \centering
    \includegraphics[width=1\linewidth]{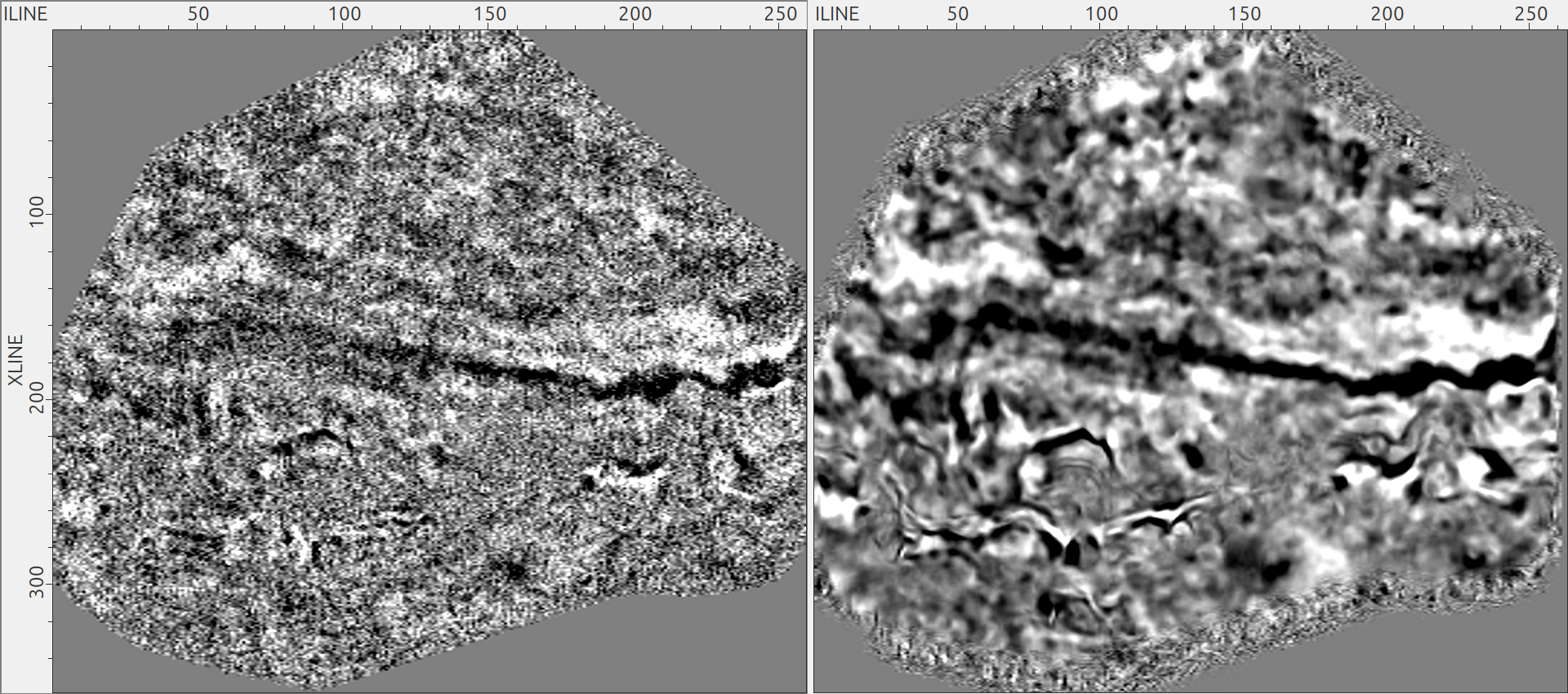}
    \includegraphics[width=1\linewidth]{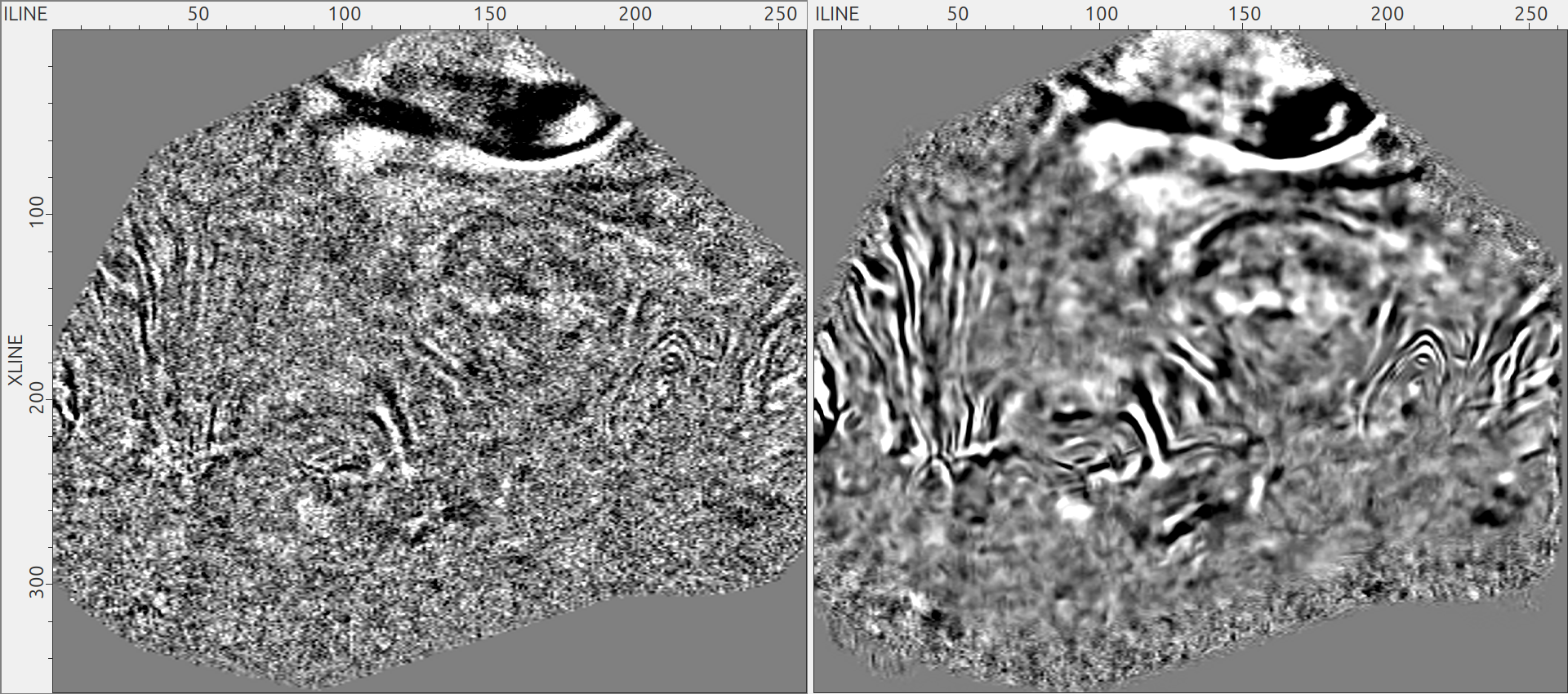}
    \includegraphics[width=1\linewidth]{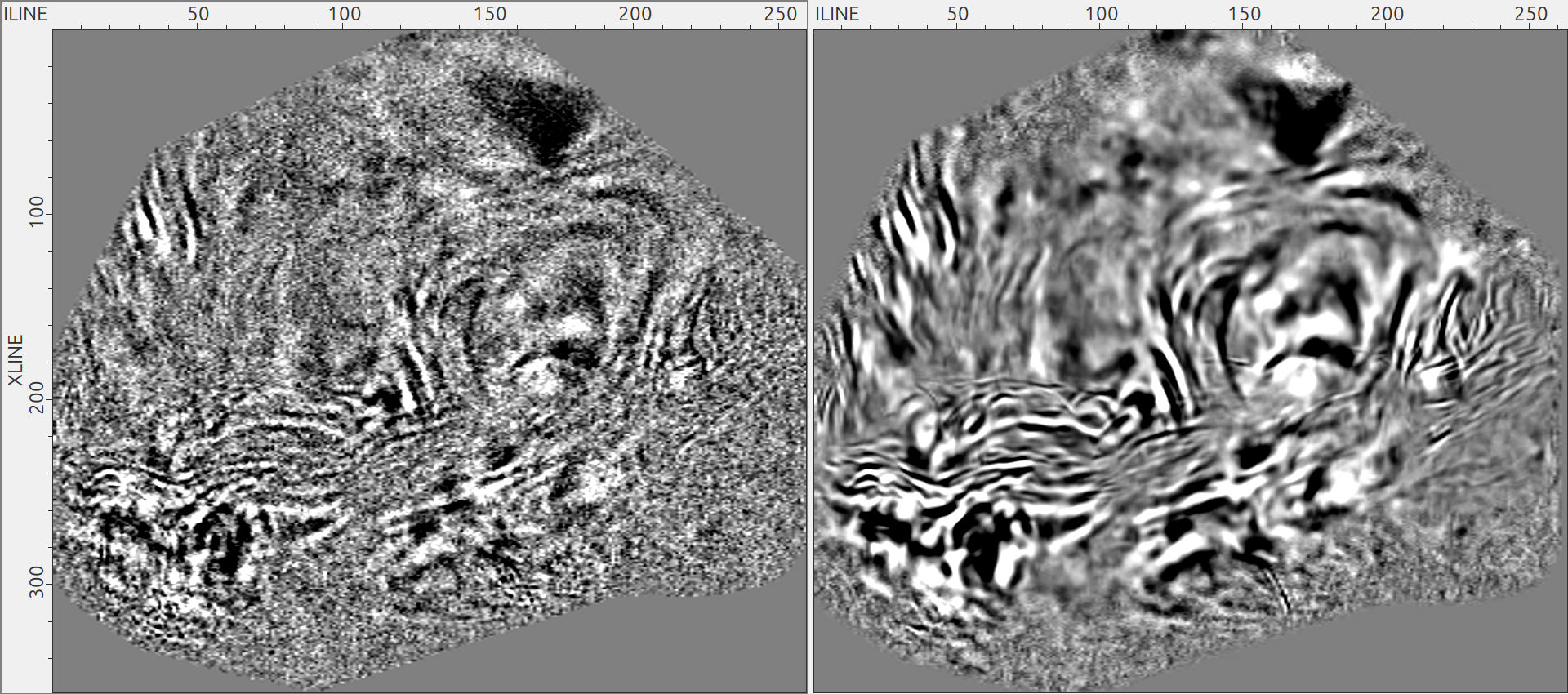}
    \caption{Comparison of time-slices at 420 (top), 724 (center), and 1048 ms (bottom) from stacked sections of the original dataset (left) and the OCT regularization (right).}
    \label{fig:ex2-ts}
\end{figure}

\begin{figure}[htbp]
    \centering
    \includegraphics[width=1\linewidth]{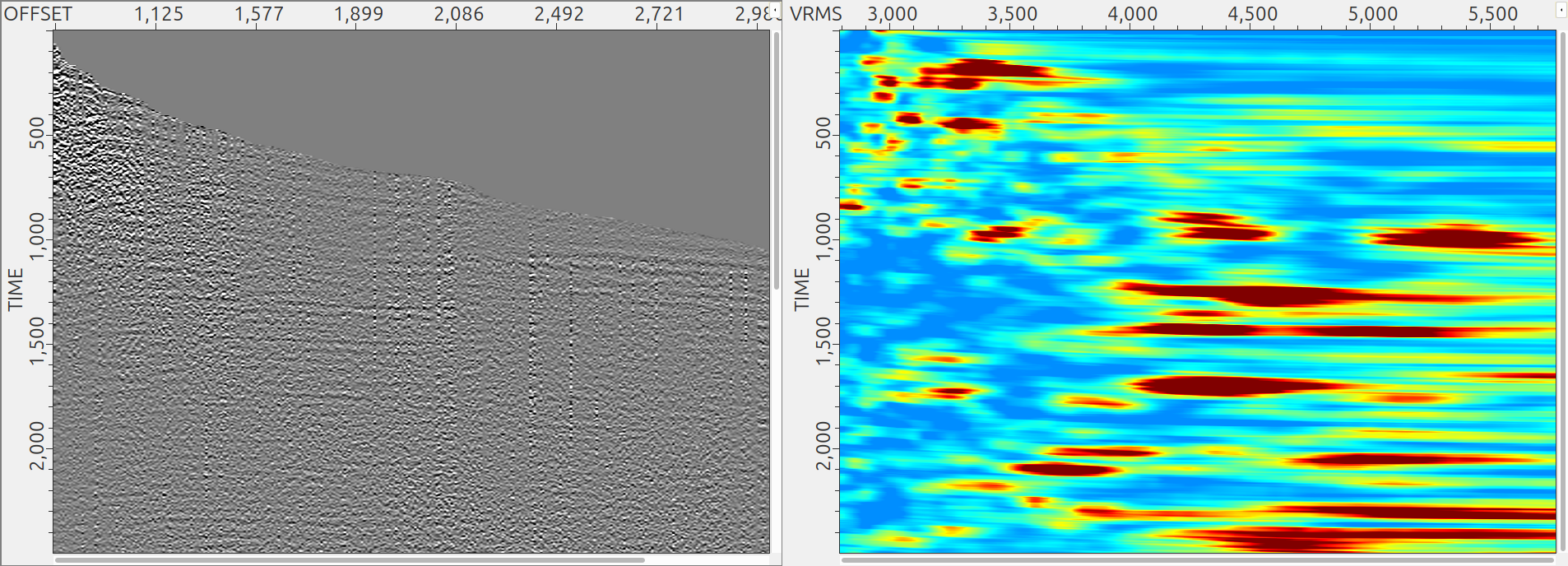}
    \includegraphics[width=1\linewidth]{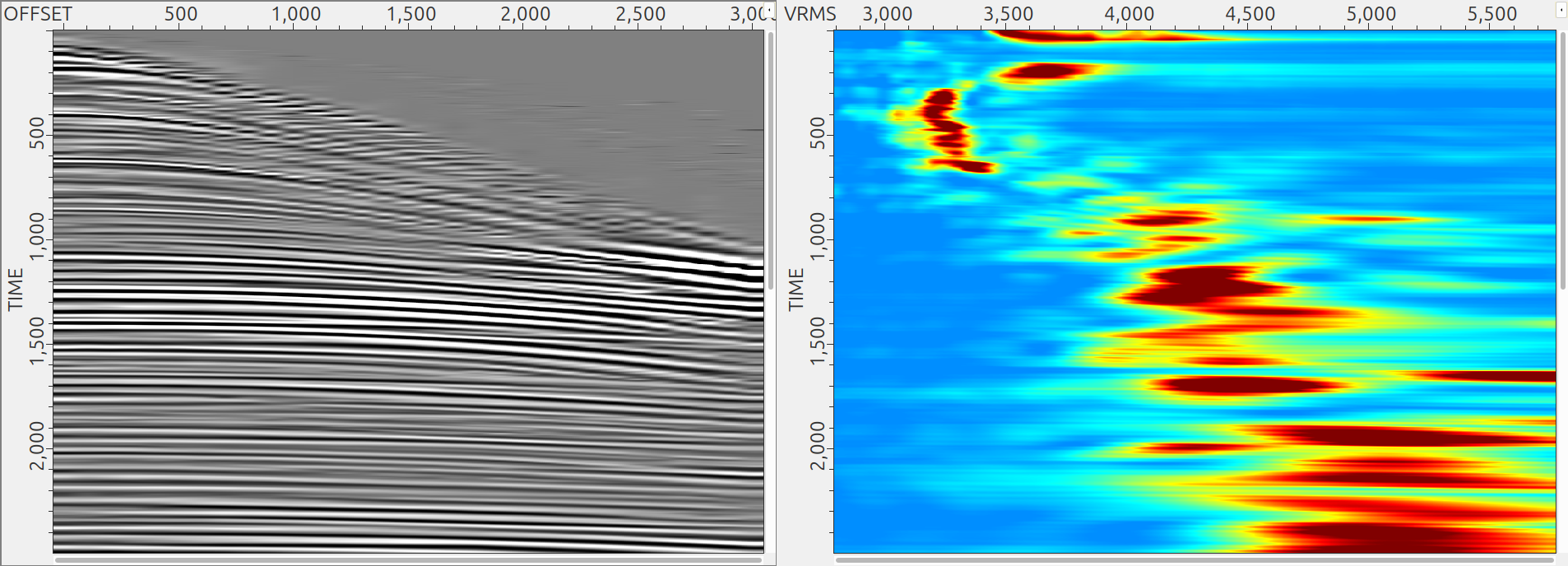}
    \caption{Influence of OCT regularization on velocity analysis for traces near the azimuth 0°. Top: Original CMP gather and its respective velocity spectrum, showing blurred energy clusters due to low signal-to-noise ratio. Bottom: Regularized gather and spectrum, where well-defined semblance maxima are clearly visible.}
    \label{fig:ex2-vel-0}
\end{figure}

\begin{figure}[htbp]
    \centering
    \includegraphics[width=1\linewidth]{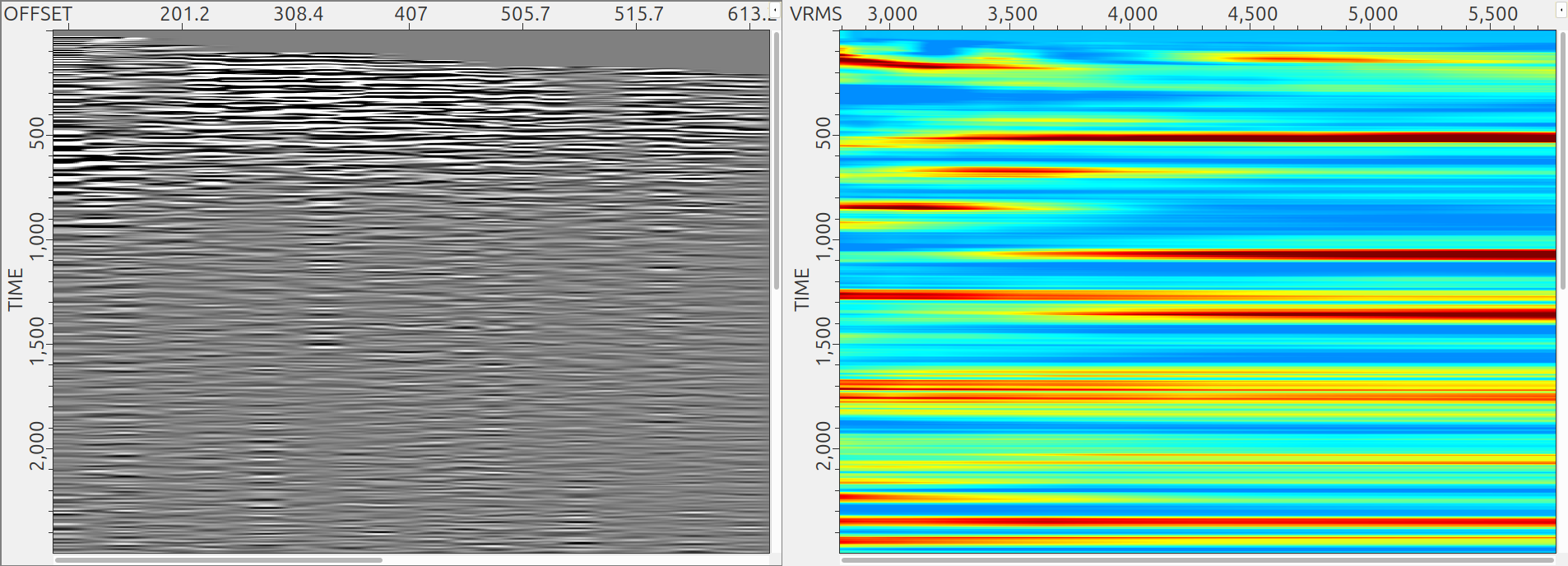}
    \includegraphics[width=1\linewidth]{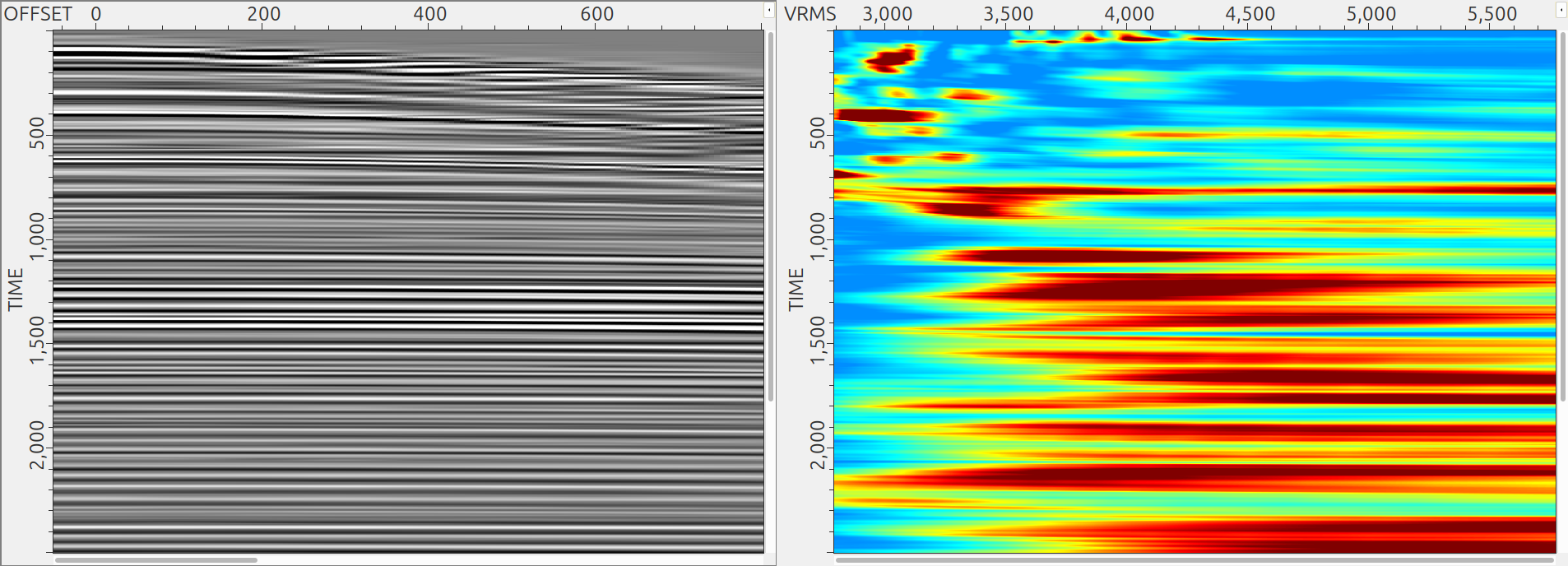}
    \caption{Influence of OCT regularization on velocity analysis for traces near the azimuth 90°. Top: Original CMP gather and its respective velocity spectrum, showing blurred energy clusters due to low signal-to-noise ratio. Bottom: Regularized gather and spectrum, where well-defined semblance maxima are clearly visible.}
    \label{fig:ex2-vel-90}
\end{figure}

Figure~\ref{fig:ex2-stks} compares stacked sections obtained from the original and regularized datasets. To ensure an unbiased comparison, both sections were stacked using identical velocity models and muting parameters. The regularized output exhibits significant noise suppression and enhanced reflector continuity. Figure~\ref{fig:ex2-ts} offers a broader view of the results, displaying three time slices that compare the original dataset with the OCT‑regularized ZO section output. This comparison reveals enhanced spatial continuity, enabling a more detailed interpretation of the events. To illustrate improvements in SNR and event continuity, we compare the velocity spectra of the regularized and original datasets. Figures~\ref{fig:ex2-vel-0} and~\ref{fig:ex2-vel-90} display the CMP gathers at 0° and 90° azimuths, respectively, along with their respective velocity spectra. The velocity trend is more precise and better defined in the regularized dataset, and it also exhibits subtle velocity variations across different azimuths.

\section{Discussion}

The presented examples show that the 5D extension of the OCT method introduced by \cite{coimbra:2016} retains its main strengths in stacking performance, enhancement, and dataset regularization. In particular, the method reconstructs coherent events and stabilizes irregularly sampled prestack data using physically consistent CRP-based kinematic trajectories. As expected, medium heterogeneity primarily controls the shape and reliability of CRP-type trajectories. When lateral velocity variations remain moderate, the proposed formulation delivers stable, accurate results provided that it uses appropriate parameter estimation and stacking apertures.

By construction, the main limitation of the approach arises in scenarios with extreme lateral velocity contrasts in the immediate vicinity of the CRP, where abrupt wavefront distortions can violate the locally smooth kinematic assumptions of the OCT formulation. In these situations, estimating CRP trajectories may become unstable. However, such extreme conditions rarely occur within small CRP neighborhoods in realistic geological settings. As a result, this limitation does not significantly affect the method's practical applicability. Fortunately, the results demonstrate that the proposed OCT framework remains sufficiently robust for real-world datasets that challenge conventional regularization and interpolation approaches.

As discussed in the main text, the numerical experiments adopt the traveltime approximation given by equation~(\ref{tCRP2}) and assume instead of symmetry of the curvature-matrix parameter operator $\mathbf{S}_0$ by $4{\bf I}/V_0^2$. Therefore, by making this assumption, the method neglects directional variations in wave-propagation velocity and reduces the number of estimated kinematic parameters to three. However, the experiments could use the more general formulation given by equation~(\ref{eq:tCRP}), which involves five parameters and enables full five-dimensional CRP-based regularization. The results clearly show that the velocity models considered in this study do not require this additional level of complexity.

The simplified parameterization strikes an effective balance between kinematic accuracy, numerical stability, and computational efficiency. In the tested scenarios, the reduced formulation captures the dominant moveout behavior and delivers high-quality reconstruction results without introducing unnecessary degrees of freedom. These findings indicate that, for a broad range of practical applications, our Ansatz approximation offers a suitable and efficient choice. At the same time, the method can reserve the full parameterization for cases involving stronger anisotropies or more complex directional velocity effects.

\section{Conclusions}

We present a physics-informed five-dimensional OCT-based stacking and regularization framework grounded in CRP kinematics. By extending the original OCT formulation to the entire prestack domain, the proposed approach addresses challenges posed by irregular sampling, noise contamination, and incomplete spatial coverage while preserving physically consistent traveltime behavior.

Applications to synthetic and field datasets demonstrate that the method improves signal-to-noise ratio, enhances structural and event continuity, and stabilizes amplitude behavior across the offset domain. By reconstructing missing traces along physically consistent CRP-based trajectories, the approach successfully recovers unrecorded information without introducing artificial events, yielding prestack volumes that are suitable for subsequent imaging and inversion workflows.

The results further show that a reduced symmetric parameterization of the CRP traveltime operator provides an effective balance between kinematic accuracy, numerical stability, and computational efficiency for a broad class of velocity models. While more general parameterizations remain available for scenarios involving more substantial anisotropies or directional velocity effects, the simplified formulation proves sufficient for the cases investigated. Overall, the proposed OCT framework constitutes a robust and practical alternative to purely mathematical interpolation techniques, offering improved data fidelity and geological consistency in complex acquisition scenarios.

\section*{Acknowledgments}

The authors gratefully acknowledge the support provided by: Petrobras (Petróleo Brasileiro S.A.) for financing this research; Shearwater GeoServices for the provision of the Academic Licence of Reveal and support from the High-Performance Geophysics Laboratory (HPG Lab); and Centro de Estudos de Energia e Petróleo (CEPETRO) at the Universidade Estadual de Campinas (UNICAMP) for their ongoing assistance throughout this research.

\bibliographystyle{elsarticle-num}
\bibliography{paper.bib}


%
\end{document}